\documentclass[
10pt, reprint,
 amsmath,amssymb,
 aps,prx
]{revtex4-2}

\usepackage{graphicx}% Include figure files
\usepackage{dcolumn}% Align table columns on decimal point
\usepackage{bm}% bold math
\usepackage{xcolor}
\usepackage{subcaption}
\usepackage{physics}
\usepackage{ragged2e}
\usepackage{mwe}
\usepackage{tcolorbox}
\usepackage{bbold}
\usepackage{tkz-euclide}
\usepackage{wasysym}
\usepackage[T1]{fontenc}
\usepackage{blindtext, rotating}
\DeclareCaptionJustification{justified}{\justifying}
\renewcommand{\vec}[1]{{\boldsymbol #1}}
\newcommand{\new}[1]{{\color{black} #1}}
\usepackage[a4paper,
bindingoffset=0.2in,
left=0.5in,
right=0.5in,
top=1in,
bottom=1in,
footskip=.25in]{geometry}
\begin{document}

\preprint{APS/123-QED}

\title{Gauge Field Dynamics in Multilayer Kitaev Spin Liquids}% Force line breaks with \\

\author{Aprem P. Joy}
\email{ajoy@uni-koeln.de}
\author{Achim Rosch}%

\affiliation{%
	Institute for Theoretical Physics, University of Cologne, Cologne, Germany
}%
\date{\today}% It is always \today, today,

\begin{abstract}
The Kitaev spin liquid realizes an emergent static $\mathbb{Z}_2$ gauge field with vison excitations coupled to Majorana fermions. We consider Kitaev models stacked on top of each other, weakly coupled by Heisenberg interaction $\propto J_\perp$. This inter-layer coupling breaks the integrability of the model and makes the gauge fields dynamic. \new{Conservation laws and topology keeps single visons immobile. However, an inter-layer vison pairs can hop with a hopping amplitude linear in $J_\perp$ confined to the layer, but their motion is strongly influenced by the type of stacking. For AA stacking, an interlayer pair has a two-dimensional motion but for the AB or ABC stacking, sheet conservation laws restrict its motion to a one-dimensional channel within the plane. For all stackings, an intra-layer vison-pair is constrained to move out-of-plane only.}
Depending on the anisotropy of the Kitaev couplings $K_x, K_y, K_z$, the intra-layer vison pairs can display either coherent tunnelling or purely incoherent hopping.
%show completely different dynamical behaviours. While coherent inter-layer tunnelling is possible for sufficiently  strong anisotropies, only incoherent processes are possible in the isotropic case.
When a magnetic field opens a gap for Majorana fermions, there exist two types of intra-layer vison pairs - a bosonic and a fermionic one. Only the bosonic pair obtains a hopping rate linear in $J_\perp$. We use our results to identify the leading instabilities of the spin liquid phase induced by the inter-layer coupling.

\end{abstract}

\keywords{emergent gauge theory, quantum spin liquids, topolgogical phase}%Use showkeys class option if keyword
\maketitle

%\tableofcontents
\section{\label{sec:intro}INTRODUCTION}

One of the most remarkable insights in modern condensed matter physics is that gauge theories may naturally emerge in the description of solid state systems \citep{visonSenthil,ZhouQSL,savaryBalents}. These gauge theories can arise from simple models of, e.g., quantum spins, and give rise to quasi-particles that exhibit fascinating properties. These include magnetic monopoles, fractional charges, Majorana fermions and anyons that are neither fermions nor bosons. Despite several theoretical models and extensive experimental research, a smoking-gun signature of emergent dynamical gauge fields in real materials is still lacking.

Among the theoretical models, perhaps the most appealing is the Kitaev honeycomb model~\citep{Kitaev06}. This two dimensional exactly solvable spin model realizes a quantum liquid, the Kitaev spin liquid (KSL), where the spins fractionalize into freely propagating Majorana fermions coupled to a static $\mathbb Z_2$ gauge field. The model also boasts a rich phase diagram featuring a gapless phase with Dirac fermions and a gapped phase with Abelian anyons. In the presence of a magnetic field one obtains a chiral spin liquid where the gauge field excitations are Ising anyons and  gapless Majorana modes emerge at the boundary of the sample.

Remarkably, the list of candidate materials that are approximately described by Kitaev models has been growing since the seminal work of Jackeli and Khaliuillin~\citep{JackeliKhaliullin}. However, non-Kitaev interactions are inevitable in these materials and as a consequence, many of them show magnetic long-ranged order. 
Here, however, an intriguing possibility is that magnetic fields may destroy the magnetic order and induce a spin-liquid state \citep{fieldinduced,nairo3}. The smoking-gun
signature of this physics, a fractional thermal quantum Hall effect has been reported in $\alpha-$RuCl$_3$ \citep{kasahara1,kasahara2}, which can be explained by the coupling of Majorana edge modes to phonons \citep{RoschQuantumHall,Balents}. Due to the strong sample-dependence of this effect, it however remains a controversial result \citep{PhysRevB.102.220404}.
\begin{figure}[h!]
	\centering
	\includegraphics[width=\linewidth]{ 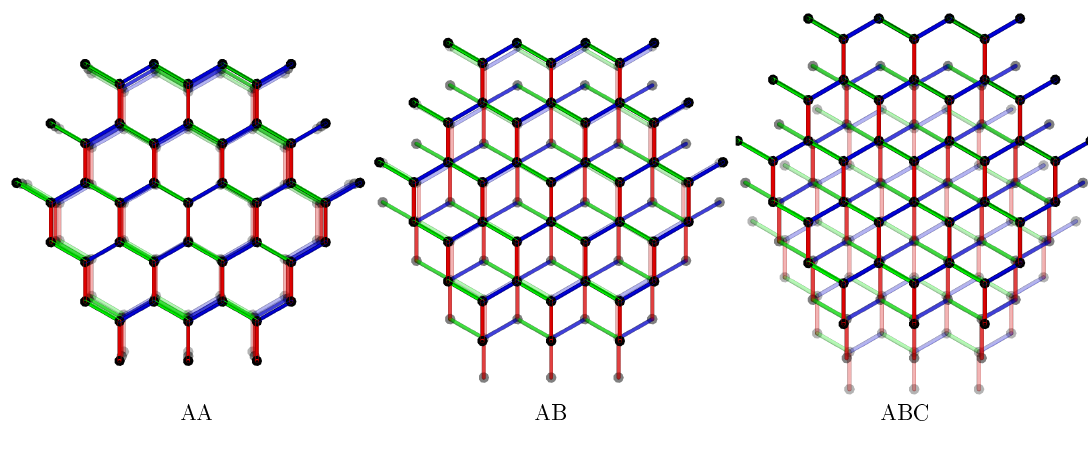}
	\caption{\small\new{\textbf{Stacking Kitaev spin liquid layers.} Top view of multilayer systems of Kitaev spin models stacked on top of each other in an AA, AB and ABC fashion (darker shaded layers lie above the lighter shaded ones). For AB stacking every second layer is shifted by $ (0,a)$ where $a$ is the nearest-neighbour distance. For ABC stacking, each layer is shifted by $(0,a)$ relative to the one below it which results in a three-layer periodic structure in the stacking direction as $(0,3a)$ is a lattice vector. The spins that sit directly above (or below) each other interact with a weak Heisenberg interlayer coupling. The coloured bonds denote the type of Kitaev interactions between the sites ($x$-blue, $y$-green and $z$-red).}}
	\label{fig:model}
\end{figure}
When studying the effects of weak perturbations to the idealized Kitaev model \citep{Songprl,KnolleAugmented18,lunkin2019perturbed,Tikhonovprl,mandal11,inti}, one of the most  important consequence of  integrability-breaking perturbations is that the formerly static gauge field becomes dynamical. In a previous publication, Ref.~\citep{prx_vison}, focusing on a single layer model, we have shown how the vison, the flux excitation of the gauge field, becomes mobile. It obtains a universal mobility at low temperatures driven by the singular scattering of Majorana modes from the vison \citep{prx_vison}. Another important type of dynamical excitations are vison pairs, studied by Zhang {\it et al.}~\citep{Batista}. Depending on the sign of the Kitaev coupling and the type of perturbation, there are regimes where vison pairs can move more efficiently than single visons \citep{prx_vison,Batista}.

An important aspect of Kitaev materials, often ignored in theoretical studies, is their three dimensional structure. Most of the candidate materials like RuCl$_3$ and certain iridates (eg. H$_3$LiIr$_2$O$_6$) consist of honeycomb lattices of magnetic atoms stacked on top. Although the inter-layer coupling is weaker than the intra-layer interactions, several experimental and theoretical studies have pointed out that it can strongly influence the magnetic and thermodynamic properties. \citep{PhysRevB.103.174417,PhysRevB.45.7430,PhysRevB.93.155143,slagle18} It was also observed that, in $\alpha-$RuCl$_3$, the magnetic ordering temperature almost doubles when stacking faults are introduced~\citep{PhysRevB.93.134423}. Interestingly, this also seems to affect the observability of the half-integer quantized thermal Hall effect discussed above~\citep{PhysRevB.102.220404}. It is therefore an important question what effects the extra spatial dimension and the inter-layer coupling have on the fractional quasiparticles that emerge from the strictly two-dimensional model. 
For example, the coupling can give rise to a pair tunnelling of Majorana fermions. How this affects inter-layer thermal transport has been studied by Werman {\it et al.}~\citep{WermanPRX}. 
\begin{table*}
	\centering
	\includegraphics[width=1\linewidth]{ 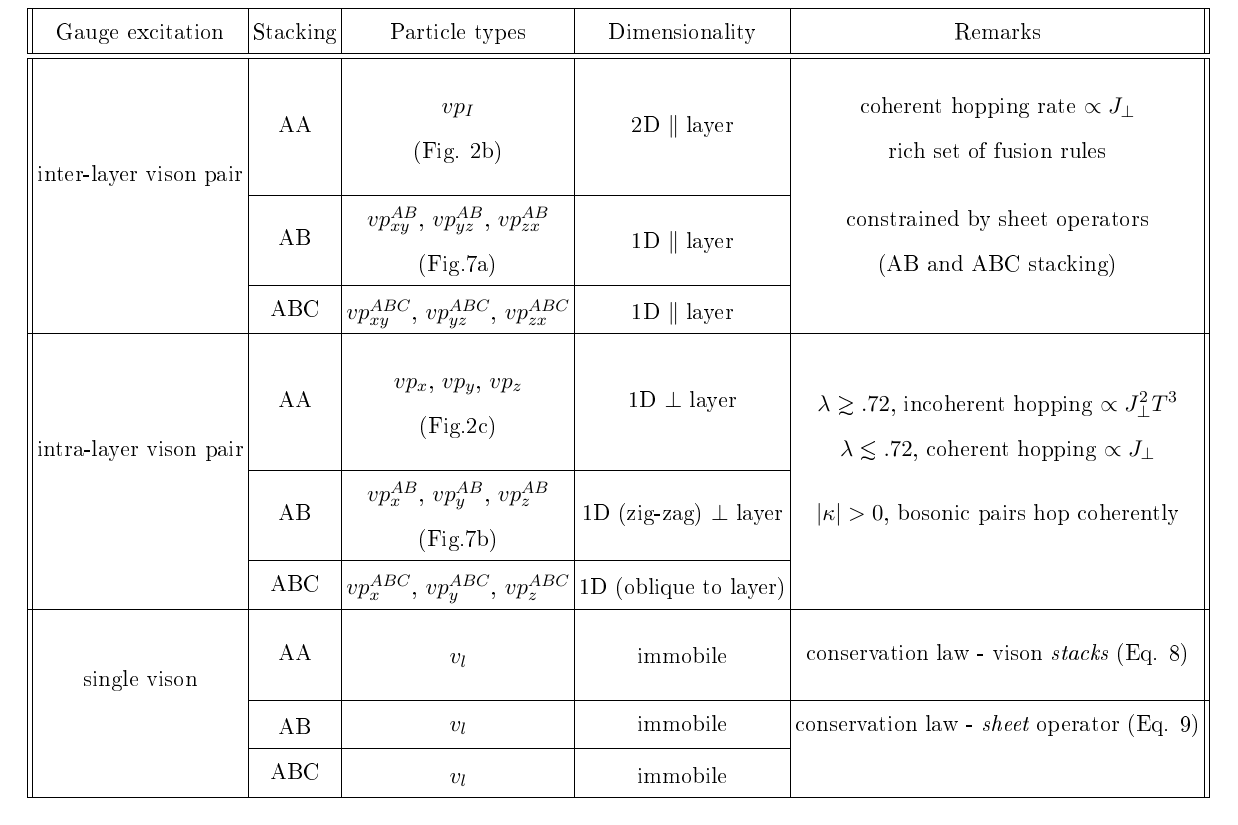}
	\caption{\new{List of the low-energy gauge excitations for multilayer Kitaev models with either AA, AB, or ABC stacking and summary of the main results. Here $\lambda = K_x/K_z=K_y/K_z$ is the Kitaev coupling anisotropy, $J_\perp$ is the strength of interlayer Heisenberg interaction and $\kappa$ is the Majorana mass gap.}}
	\label{tab:summary}
\end{table*}
In this context, recent studies on bilayer Kitaev models have provided some insights into the stability and phase transitions of the pure Kitaev phase \citep{Tomishige19,Urbanbilayer1,Urbanbilayer2}. For example, exact diagonalization and mean-field studies on bilayer Kitaev models have shown that the signs of the Kitaev coupling in the two layers do not affect the results as long as they are the same (FM or AFM) whereas an entirely different phase diagram is obtained when an FM layer is coupled to an AFM layer \citep{Urbanbilayer2}. 

It was shown in Ref.~\citep{tcbilayer}
that in a bilayer toric-code model, an  Ising inter-layer interaction can trigger a phase transition from one to another topological phase. Coupled Toric code layers have also been used as starting points for constructing several exotic 3D phases including fractonic ones \citep{Ma17,slagle17,KLayerToricCode}.
Both from a theoretical and experimental  point of view, it is therefore important to understand the properties of  emergent excitations of the Kitaev model when multiple layers are weakly coupled in a three dimensional system. Our goal is therefore to investigate the dynamics of gauge-field excitations in a simple model of Kitaev layers coupled by a Heisenberg interaction. After introducing the model, we identify vison pairs as the dominant mobile excitations and calculate their dynamical properties. 
\section{RESULTS}

\subsection{Model and Hamiltonian}
\begin{figure}[t]
	\centering
	\includegraphics[width=\linewidth]{ 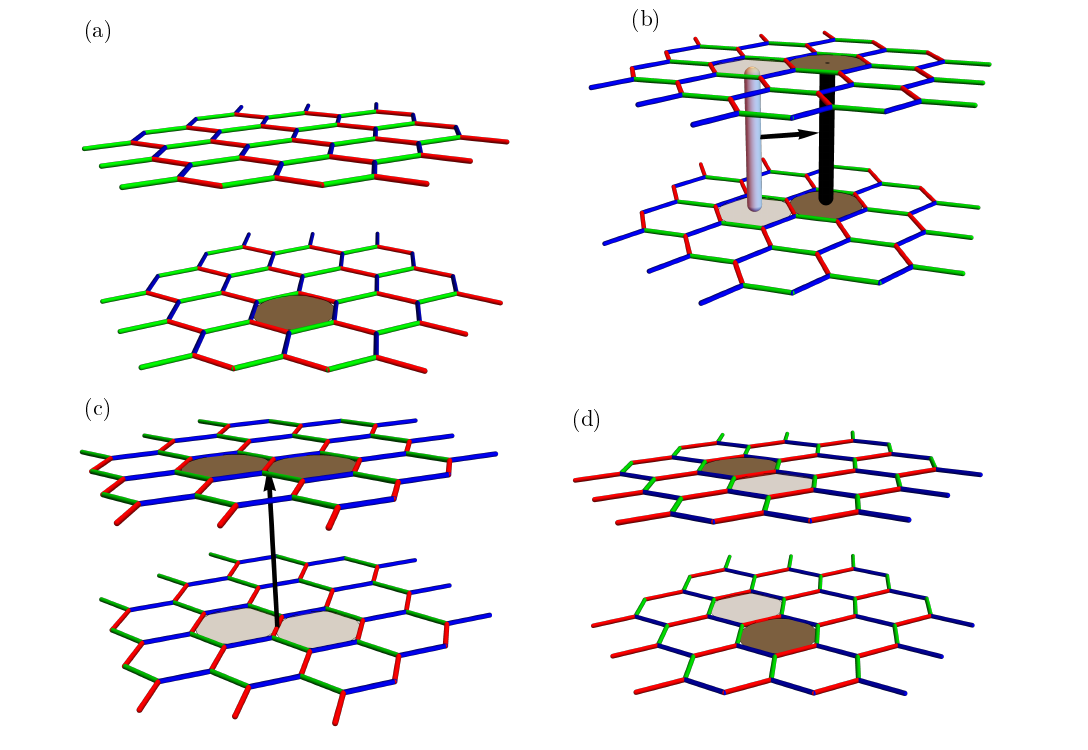}
	\caption{\small {\bf \new{Visons in an AA stacked Kitaev model}.} (a) A single vison (shaded plaquette) excitation is immobile. (b) An inter-layer vison pair $vp_{I}$ can hop in a plane parallel to the layers. (c) An intra-layer vison pair $vp_\alpha$ ($\alpha=x,~y,~z$)  can only hop between layers along the stacking direction. (d) Another inter-layer pair $vp'_{\alpha}$, flips between two states by exchanging the vison positions. In (b), (c) and (d), the inter-layer coupling induces transitions between the light- and dark-shaded plaquette pairs.}
	\label{fig:illustration}
\end{figure}

We consider multiple stacked layers of Kitaev models coupled by a weak inter-layer Heisenberg interaction. \new{In this work, we focus on three main types of stacking patterns  AA, AB and ABC, as illustrated in Fig.~\ref{fig:model}. In honeycomb materials like $\alpha$-RuCl$_3$, either AB or ABC stacking (or a mixture of both) is usually observed at low temperatures\cite{PhysRevB.93.134423}. For AA stacking, adjacent layers are related by translation by an interlayer distance $d_\perp$ in the $z$ direction, perpendicular to the plane. For AB stacking, however, adjacent layers obtain an extra in-plane shift, such that atoms on the A sublattice in even layers and atoms of the B sublattice in odd layers are on top of each other. Finally, adjacent layers are all shifted by a lattice constant relative to each other in the same direction for ABC stacking, see Fig.~\ref{fig:model}.} The Hamiltonian for our model is therefore,
\begin{align}
H=\sum_{l} H^l_{K}+\Delta H_\perp 
\label{eq:hamiltonian}
\end{align}  
where
\begin{align}
H^l_{K}=\sum_{\langle ij\rangle_\alpha}K_\alpha \sigma^\alpha_{i,l}\sigma^\alpha_{j,l},
\end{align}
with $\alpha=x,y,z$ defined as in Fig.~\ref{fig:illustration}a, is the single layer Kitaev Hamiltonian for layer $l$. \new{We assume $K_x=K_y=K_z=-1$ for our calculations in the isotropic limit.
	The form of the interlayer coupling term depends on the stacking. For the AA model,
	\begin{align}
	\Delta H^{AA}_\perp=J_\perp \sum_{l}\sum_{\langle ij \rangle_\perp}\vec{\sigma}_{i,l}\cdot \vec{\sigma}_{j,l+1} .
	\end{align}
	For an AB stacked model,
	\begin{align}
	\Delta H^{AB}_\perp=J_\perp \sum_{l}\sum_{
		\substack{
		\langle ij \rangle_\perp \\ 
		i \in A, j \in B
		%}
	}}
	\vec{\sigma}_{i,2l}\cdot (\vec{\sigma}_{j,2l+1}+\vec{\sigma}_{j,2l-1})
	\end{align}
	and for ABC stacking
	\begin{align}
	\Delta H^{ABC}_\perp=J_\perp \sum_{l}\sum_{
		\substack{
			\langle ij \rangle_\perp \\ 
			i \in A, j \in B
			%}
	}}
	\vec{\sigma}_{i,l}\cdot \vec{\sigma}_{j,l+1}
\end{align}
	where $\langle ij \rangle_\perp$ denotes nearest neighbour sites separated by interlayer spacing $d_\perp$ along the stacking axis direction, see Fig.\ref{fig:illustration}. We note that we include only interactions between spins that lie directly above (or below) each other. Therefore, for AB stacking, only the spins in even layers on the A sublattice interact with B-sublattice spins in odd layers. Similarly, for ABC stacking, a spin in layer $l$
on the A sublattice	interacts with a spin in layer $l+1$ on the B sublattice directly above its position. In contrast, a spin in layer $l$
on the B sublattice	interacts with a spin in layer $l-1$ on the A sublattice directly below its position.}

For $J_\perp = 0$, the model reduces to decoupled layers of the exactly solvable pure Kitaev model~\citep{Kitaev06}. The solution is given by expressing the spin operators in terms of four species of Majorana fermions $b^x,b^y,b^z,c$. 
\begin{align}
\sigma^\alpha_{i,l}=i b^\alpha_{i,l}c_{i,l}
\end{align}
Defining link operators $u^l_{\langle ij \rangle_\alpha}=i b^\alpha_{i,l}b^\alpha_{j,l}$ that commute with the pure Kitaev Hamiltonian $H_K$, we can write
\begin{align}
H^l_{K} = \sum_{\langle ij \rangle_\alpha} K_\alpha u^l_{\langle ij \rangle_\alpha}ic_{i,l}c_{j,l}
\label{eq:majorana_Ham}
\end{align}
The link operators have eigenvalues $\pm 1$ and realize a $\mathbb{Z}_2$ gauge field which is coupled to the itinerant Majorana fermions.
There is an extensive number of conserved quantities defined by the plaquette operators 
\begin{align}
\hat{W}_{p,l}=\prod\limits_{\hexagon} \sigma_{i,l}^\gamma \sigma_{j,l}^\gamma=\prod\limits_{\hexagon} u^l_{\langle ij \rangle_\alpha}
\end{align} 
with eigenvalues $W_{p,l}=\pm 1$, where $p$ labels the hexagonal plaquette in the layer $l$. These operators can be identified with objects that live on the hexagons and constitute the physical degree of freedom of the gauge field, {\em visons}. A plaquette $p$ in layer $l$ is said to host a vison if $W_{p,l}=-1$. For $J_\perp=0$, the visons are fully localized and have a finite energy $E^0_v\approxeq 0.15 |K|$. The ground state is hence free of visons and gauge fixing all $u^l_{\langle ij \rangle_\alpha}=1$ leads to a Dirac spectrum for the Majorana fermions with $\omega(\vec k)=\frac{\sqrt{3}|K|}{2} k$. A vison acts as a singular Aharanov-Bohm potential for the Majoranas as it carries a  gauge flux of $\pi$.

\subsection{\new{Conservation Laws and Dynamics}}

We now identify and study the dynamical excitations in a multilayer system for finite but small $J_\perp$.
For sufficiently small $J_\perp$, the Kitaev phase remains stable. As visons and Majoranas within one layer can only be created in pairs, a single vison or a single Majorana is constrained to move within each plane \citep{Kitaev06,Urbanbilayer1}. 

\begin{figure}
	\centering
	\includegraphics[width=\linewidth]{ 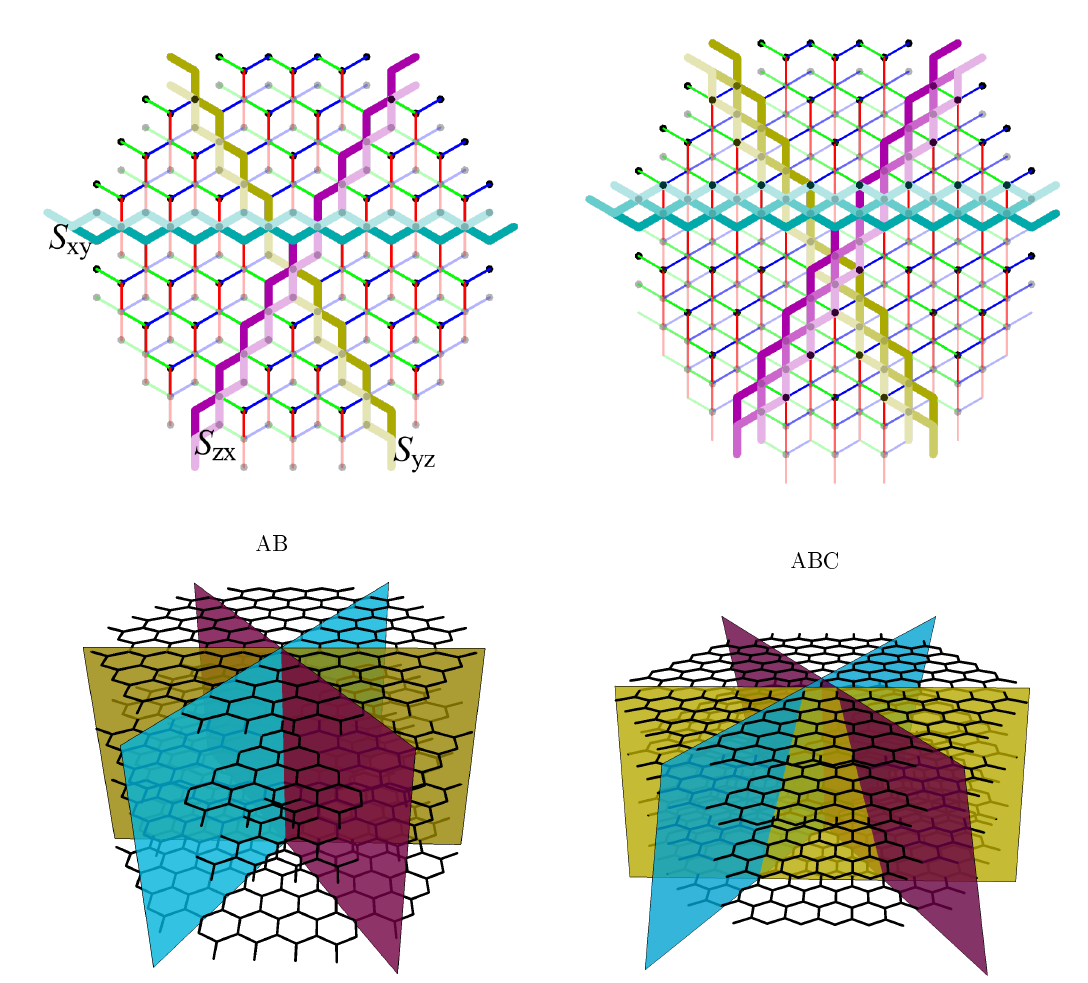}
	\caption{\new{\small Conserved sheet operators in AB (left) and ABC (right) stacked Kitaev models shown for a bilayer and trilayer respectively (darker shaded layers are above the lighter shaded ones.). A single sheet operator is defined as the product of Kitaev terms along a thick colored chain. This defines three types of sheets formed by $x$ and $y$ bonds (cyan), $y$ and $z$ bonds (yellow), and, $x$ and $z$ bonds (violet). The full 3D representations for the sheet operators are shown in the lower panels. For ABC stacking the sheets are tilted.}}
	\label{fig:ABloops}
\end{figure}
\new{However, remarkably, due to the presence of residual conserved quantities, a single vison stays immobile in AA, AB and ABC stacked models. 
	In the AA model, it was observed in Ref.~\citep{Tomishige19} and \cite{Urbanbilayer1} that there exists a conserved quantity, the {\em vison stack} operator, defined as a product of  vison operators located on top of each other
	\begin{align}
	X_p =\prod_{l} W_{p,l}.\label{eq:visonStackOp}
	\end{align} $[X_p,H]=0$ implies that the parity of visons in a given column
	above the plaquette $p$ is conserved.
	While this prevents the in-plane motion of a single vison, the out of plane motion is forbidden due to the total vison parity conservation within a layer (this vison parity is the product of all plaquettes in a layer which evaluates to the identity operator for periodic boundary conditions).
	Thus it remains completely localized.
	The $X_p$ commute with each other and  the total number of commuting conservation laws scales therefore with $2^{L^2}$, where $L$ is the linear (in-plane) dimension of the system.

	In the AB and ABC stacked models, such a conserved stack operator does not exist. However, we identify a different set of global conserved quantities that strongly restricts the dynamics of flux excitations and, remarkably, keep a single vison immobile. 
	We define conserved {\em sheet} operators $S^{xy}_m$, $S^{xz}_m$, and $S^{yz}_m$ as
	\begin{align}
	S^{\alpha \beta}_m = \prod_l L_{m,l}^{\alpha \beta} = \prod_l \left( \prod_{\langle ij\rangle_ \in \mathcal{C}_{m,l}^{\alpha \beta}} \hat{K}^l_{\langle ij\rangle} \right)
	\label{eq:sheets}
	\end{align}
	where $\hat{K}^{l}_{\langle ij\rangle}=\sigma_{i,l}^\alpha \sigma_{j,l}^\alpha$ are the Kitaev terms on the bond $\langle ij \rangle$ and $\mathcal{C}_{m,l}^{\alpha \beta}$ is a chain made of $\alpha$ and $\beta$ bonds in layer $l$ that wraps around the torus (periodic boundary conditions are assumed) as shown in Fig.~\ref{fig:ABloops}. Here $m=1,\dots, L$ denotes which chain in a given layer has to be chosen.
Chains in neighboring layers are parallel to each other,  see Fig. \ref{fig:ABloops}.  The collection of the chains defines a sheet. For AB stacking the sheet is always perpendicular to the layer (or, more precisely, form a zig-zag structure in $z$ direction). In contrast, the sheet is tilted for ABC stacking, see  Fig. \ref{fig:ABloops}.
	 %Specifically, we use Wilson loops  $L^l_{xy}$ and $L^l_{xz}$ that span the two linearly independent directions on the torus. 
	As explicitly shown in Eq. \eqref{eq:sheets}, the sheet operators are simply the product of Wilson loops $ L_{m,1}^{\alpha \beta}$ of each layer. 
	%Top panel of Fig. \ref{fig:response} shows the full 3D structure of the sheet operators for AA, AB and ABC stacked models.
	The operators $S^{xy}_m$, $S^{yz}_m$ and $S^{xz}_m$ do not commute with each other but the product $S^{\alpha\beta}_{m_1}S^{\alpha\beta}_{m_2}$ commutes with all other operators. For AB and ABC stackings, we find that the number of commuting operators scales with $\frac{2^{3 L}}{8}$. Thus, the number of commuting conservation laws is lower compared to the AA stacking case where we found it to be $2^{L^2}$. Nevertheless, they put very strong constraints on the dynamics of gauge excitations as specified below.

	%\prod_{\vec r \in l_1}\sigma^x_{1,\vec r}\sigma^x_{1,\vec r+\vec dx} \sigma^y_{1,\vec r}\sigma^y_{2,\vec r-\vec dy}\sigma^x_{1,\vec r}\sigma^x_{1,\vec r-\vec dx} \sigma^y_{1,\vec r}\sigma^y_{1,\vec r+\vec dy}

	Let us now consider an AB-stacked bilayer and examine the motion of a single vison excitation. To hop a single vison at position $\vec R$ in layer 1, to $\vec R'$ within the layer, one must apply an open string (product) of spin operators $C^1$ with its ends at $\vec R$ and $\vec R'$. Within our model, strings  arise at $n^{th}$ order perturbation theory expressed as $C^1 \sim (\Delta H_{12})^n$ where $n$ is a positive integer. However, any such open string will inevitably cross at least one of the Wilson loops $ L_{m,1}^{\alpha \beta}$ flipping its parity. Due to the conservation of the corresponding sheet operator $S^{\alpha\beta}_m =  L_{m,1}^{\alpha \beta} L_{m,2}^{\alpha \beta}$, this must be accompanied by a parity flip of its partner $ L_{m,2}^{\alpha \beta}$ which results in the creation of visons in layer 2. Hence a single vison cannot move without creating extra visons in another layer! This argument also applies to the ABC model since the same conservation law is valid except for a ``tilting'' of the sheets in this case.
	Thus, in the single-vison sector, (dressed) visons are immobile both for all three stackings considered in this work.}

%\begin{figure}
%	\centering
%	\includegraphics[width=\linewidth]{ Fig9.pdf}
%	\caption{\new{\small Conserved sheet operators in a full 3D stacked multilayer for different stackings.}}
%	\label{fig:sheets3D}
%\end{figure}
This is, however, not true for vison pairs, which we will therefore study in the following.
For such pairs, we can compute the hopping matrix elements using perturbation theory. To linear order in $J_\perp$, the hopping matrix element is
\begin{align}
t = \bra{\Psi(\{\vec R_l\})}\Delta H_\perp \ket{\Psi(\{\vec R'_l\})},
\label{eq:matrix_element}
\end{align}
See methods.
Using the exact eigenstates obtained in the $J_\perp=0$ limit, one can compute this matrix element using  Pfaffian determinants \citep{robledo,prx_vison}.

\begin{figure*}[!t]
	\centering
	\includegraphics[width=\linewidth]{ 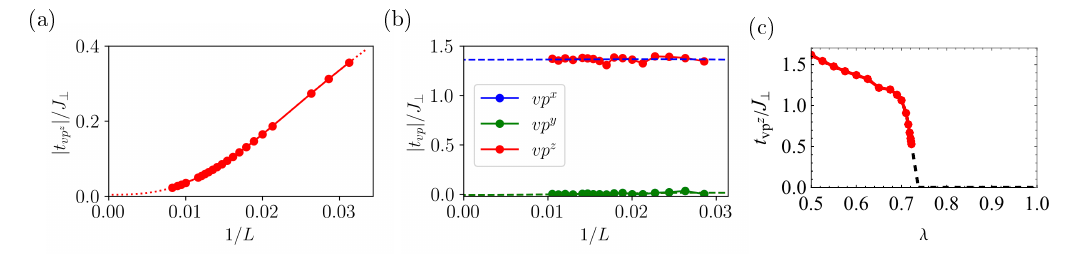}
	\caption{\small {\bf Hopping of intra-layer vison pairs \new{in the AA stacked model.}} (a) Inter-layer hopping amplitude $t_{vp_z}$ of intra-layer vison pairs, as a function of inverse system size at the isotropic point, $\lambda=1$. In the thermodynamic limit, $t_{vp_z}$ vanishes. (b) For anisotropic Kitaev couplings, $\lambda=0.6$, the vison pair $vp_z$ obtains a finite hopping amplitude while $vp_x$ and $vp_y$ do not. (c) Hopping amplitude of $vp_z$ as a function of anisotropy $\lambda$ for a system size of $L = 70$. The dashed lines are guides to the eye.}
	\label{fig:tvp2}
\end{figure*}
\subsection{Intra-layer vison pair - AA stacking}
\subsubsection{Coherent hopping}
\new{We first consider the AA stacked multilayer and 
} define an {\em intra-layer vison pair}, $vp_\alpha$, as an object with two visons on adjacent plaquettes sharing a $\alpha$ type link, $\alpha=x,y,z$. While they cannot move within the layer due to the conservation of $\hat X_p$, they may hop across layers along the stacking direction. The action of $\Delta H^\alpha_\perp$ on a $vp_\alpha$ excitation annihilates it and creates a $vp_\alpha$ pair in the adjacent layer on the $\alpha$ bond directly on top (or bottom). This results in an effective inter-layer hopping as illustrated in Fig.\ref{fig:illustration}b. 
To linear order in $J_\perp$, the inter-layer hopping matrix element of a vison pair sharing a $z$ link $\langle ij\rangle$ is
{\small \begin{align}
	\frac{t_{vp_z}}{J_\perp}=& \Big(\bra{\Phi^{2}_0(\emptyset)}\otimes \bra{\Phi^1_0(\vec R)}\Big)\Delta H_\perp \Big(\ket{\Phi^{1}_0(\emptyset)}\otimes \ket{\Phi^{2}_0(\vec R)}\Big)\nonumber \\ 
	=&\Big|\bra{\Phi^1_0(\vec R)}\sigma^z_{i,1}\ket{\Phi^{1}_0(\emptyset)}\Big|^2+\Big|\bra{\Phi^1_0(\vec R)}\sigma^z_{j,1}\ket{\Phi^{1}_0(\emptyset)}\Big|^2.
	\label{eq:vp2_hop}
	\end{align}
}

The matrix elements are evaluated numerically, see methods, for a finite size system and plotted as a function of system size in Fig.~\ref{fig:tvp2}.
For an isotropic Kitaev model, Fig.~\ref{fig:tvp2}a,
the hopping rate of the vison pair vanishes in the thermodynamic limit. Thus we find that there is in no coherent hopping linear in $J_\perp$! 
This result is directly related to previous results obtained for single-layer Kitaev models \citep{knolle,Loss,vojta}. When one compares the ground-state wave functions in the presence and absence of a vison pair, they have different matter-Majorana parity.
Thus, the only way to create or destroy a vison pair, is to simultaneously change the Majorana parity.
Technically, in the calculation of Eq.~\eqref{eq:vp2_hop} this is accomplished by occupying the lowest-energy Majorana state. In the gapless phase, this mode is delocalized and hence the tunnelling of such an extended object (vison pair + fermion) vanishes in the thermodynamic limit. While Eq.~\eqref{eq:vp2_hop} is only valid to linear order in $\Delta H_\perp$, the `parity-obstruction argument' applies to all orders in $\Delta H_\perp$ as long as the Kitaev phase is stable.

The fact that the coherent tunnelling rate of 
Eq.~\eqref{eq:vp2_hop}  vanishes in the thermodynamic limit, does, however, not imply that vison pairs
cannot move from layer to layer. At finite $T$, we will later show that the visons can move incoherently from layer to layer by scattering from 
Majorana modes.

The discussion given above is valid for approximately isotropic Kitaev couplings. The qualitative properties of the vison pair change, however,  completely, if we make the Kitaev couplings anisotropic, say by tuning $\lambda=K_x/K_z=K_y/K_z$. As shown in Fig. \ref{fig:tvp2}b and c, the vison pair $vp_z$ obtains a finite hopping matrix element in the thermodynamic limit when $\lambda \apprle 0.72$. In contrast, the other two types of vison pairs $vp_x$ and $vp_y$ are still not able to hop coherently.

Note that for $\lambda > 0.5$, the Majorana fermions are gapless and the ground-state wavefunction of the system are adiabatically connected to the isotropic point $\lambda=1$. Thus the abrupt change of the tunnelling rate at $\lambda \approx 0.72$ has to come because the wave function of the vison pair $vp_z$ changes its matter-parity at this point. We will argue below that, at the same point, the lowest energy vison pair changes from a fermionic ($\lambda\apprge 0.72$) to a bosonic excitation ($\lambda\apprle 0.72$). Only the boson can hop coherently as discussed in detail later.
This change of quantum numbers of an excited state is sometimes called a `dynamical quantum phase transition' (DQPT) as at this point, time-dependent correlation functions change their qualitative properties. 
The same DQPT also affects spin-spin correlation functions of the pure Kitaev model as has been pointed out by Knolle {\em et al.} \citep{knolle}.

\begin{figure*}
	
	\centering
	\includegraphics[width=\linewidth]{ 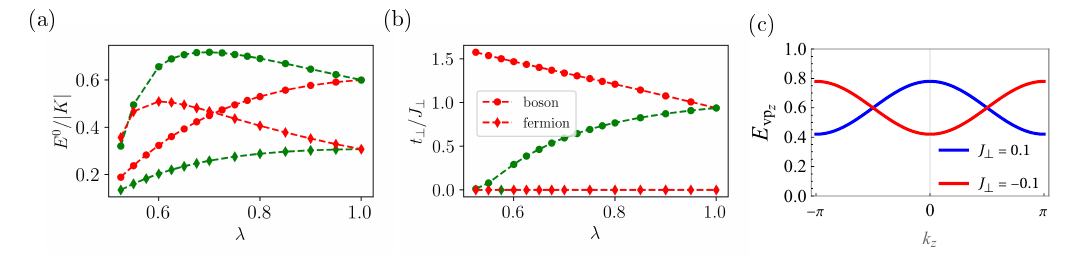}
	\caption{\small {\bf Ising anyon pairs - \new{AA stacking.}} (a) The energy of intra-layer vison pairs (or, equivalently, Ising anyon pairs) in the non-abelian phase (for $\kappa=0.1, ~J_\perp=0$) as a function of the anisotropy $\lambda$.
		The colors label the type of bond shared by the anyons - $vp_x$ and $vp_y$ vison-pairs (green) have the same energy, the $vp_z$ pair is colored in red. Circles and diamonds denote bosonic and fermionic vison pairs, respectively. A level crossing between the fermionic and bosonic vison-pairs occurs at $\lambda \approx 0.72$. (b) Inter-layer hopping amplitudes of the vison pairs. Only pairs that belong to the vacuum sector (i.e, bosons) can coherently hop. (c) The one dimensional dispersion of the bosonic $vp_z$ along the $z$ direction for $\lambda=1$ and $\kappa=0.1$ for different signs of $J_\perp$.}
	\label{fig:level_crossing}
\end{figure*}
\subsubsection{Majorana assisted hopping} 
\label{sec:majorana_assisted}
Coming back to the isotropic limit (or, more precisely, $\lambda\apprge 0.72$), we have argued that the hopping of the vison pair ($vp_\alpha$) between layers is always accompanied by the creation (or destruction) of Majorana modes in the two layers. If we denote the operator creating a dressed vison pair in layer $l$ by $p^\dagger_{i,l}$, the resulting Majorana-assisted hopping can schematically be written as $H_\text{eff} \propto\sum_{i,\langle l,l'\rangle} c_{i,l'}\,  p^\dagger_{i,l'} p_{i,l}\, c_{i,l}$. Instead of calculating $H_\text{eff}$, it is easier to compute directly the incoherent tunnelling rate of the vison pair at finite temperatures.

According to Fermi's golden rule, it is given by
\begin{align}
\Gamma^z_\perp =  \frac{2\pi}{\hbar}\sum_{m,n}&|\bra{\Psi_m(\vec R_l)}\Delta H \ket{\Psi_n(\vec R_{l+1})}|^2 \nonumber \\
& \qquad \qquad \quad \times e^{-\beta E_n} \delta(E_m-E_n)
\end{align}

The states $\ket{\Psi(\vec R_l)_m}=\ket{\Phi_{m_1}(\vec R_l)}\ket{\Phi_{m_2}(\emptyset)}$ are constructed by populating the Majorana states of layer $l$ and $l+1$ labelled by $m_l$ and $m_{l+1}$, respectively, with $m = \{m_l,m_{l+1}\}$.

In the low energy (and thus low $T$) limit, one can neglect the scattering of Majorana fermions from the vison pair. This can be confirmed by calculating the full Green's function of the Majoranas in the presence of a vison-pair which approaches the free particle Green's function as $\omega\to 0$, see Supplementary information (also Ref.~\citep{aadityaPRB}). (Note that, this is in sharp contrast to the effect of an isolated vison which results in a highly singular scattering when $\omega\to0$ \citep{prx_vison}). In this limit, the Lehmann representation approximately transforms the above expression into a convolution of two local  Majorana spectral functions $C_{\alpha \beta}(\omega,0)\propto |\omega|$ in each layer, see Supplementary information for details. This leads to an inter-layer diffusion constant for the intra-layer pairs
\begin{align}
D_\perp=\Gamma^z_\perp d_\perp^2 \sim \frac{J_\perp^2 d_\perp^2  T^3}{K^4}
\end{align}
where $d_\perp$ is the inter-layer separation.
The Majorana-assisted hopping of vison pairs is therefore possible, but strongly suppressed at low temperatures.
\subsubsection{Time reversal breaking and Anyon tunnelling}
\label{sec:non_abelian}
In the presence of an external magnetic field, the gapless KSL transform into  a chiral non-abelian phase where the matter fermions $\psi$ acquire a gap $m=6\sqrt 3\kappa$ due to a field-induced next-nearest neighbour hopping $\kappa$ \citep{Kitaev06}.
In this case, single visons become Ising anyons ($\sigma$) with localized Majorana zero modes attached to them. The intra-layer vison-pairs (anyon pairs) now carry a localized fermion mode with energy $\epsilon_0$, where $m>\epsilon_0>0$,
arising from the hybridization of the Majorana modes of the two visons.
This  mid-gap mode can be occupied or empty. In the language of topological field theories, the Ising anyons can fuse in two different ways, $\sigma \times \sigma = \mathbb{1}$ or $\sigma \times \sigma=\psi$. An anyon pair that is created by any physical operator out of vacuum is a boson whereas a pair that fuses to release a fermion ($\psi$) is fermionic in nature. The energies of these vison pairs are shown in Fig.~\ref{fig:level_crossing}a. The two types of pairs have different total fermionic parity and do not hybridize with each other. 

Fig.~\ref{fig:level_crossing}b shows the tunnelling rate of the vison pairs to linear order in $J_\perp$. 
At $T=0$, only the bosonic pair can tunnel between the layers (to arbitrary order in $J_\perp$), whereas  the fermionic pair requires assistance from thermally excited Majorana fermions which are gapped in the non-abelian phase. 

The  level crossing between the fermionic and bosonic $vp_z$ pair at  $\lambda\approx 0.72$ shown in Fig. \ref{fig:level_crossing} has important physical consequences in the $\kappa \to 0$ limit, when the gap closes. It explains our previous finding from Fig.~\ref{fig:tvp2}c: the nature of the vison-pair changes suddenly  
at this point in a so-called `dynamical quantum phase transition'.

When $\kappa$ and therefore the Majorana gap is reduced, the fermionic bound state grows in size. Thus, for $\kappa \to 0$, only the vison pair without a fermion bound to it (the lower energy state) remains as a point-like particle. Therefore, the `physical' $vp_z$ pair at $\kappa=0$ changes its statistics and ability to tunnel at $\lambda \approx 0.72$. In contrast, the  $vp_x$ and $vp_y$ pairs do not tunnel in this limit.

Note that, we expect that the critical value, $\lambda \approx 0.72$ calculated above in the $J_\perp \to 0$ limit, will be shifted to larger values upon increasing $J_\perp$ as the bosonic vison gains extra kinetic energy  by inter-layer tunnelling.

\subsection{Inter-layer vison pair - AA stacking}
\begin{figure*}
	\centering
	\includegraphics[width=\linewidth]{ 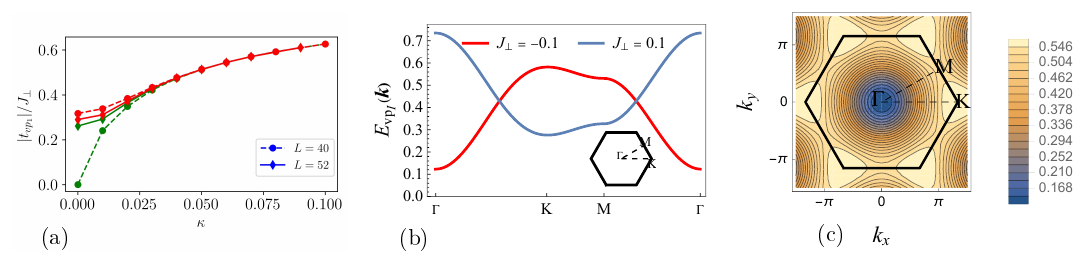}
	\caption{\small \new{\textbf{Dynamics of interlayer vison pairs - AA stacking.} In Figure (a), the hopping amplitude of the interlayer-vison pair, $vp_I$, is plotted as a function of Majorana mass $\kappa$ for the AA stacked model. The corresponding 2D dispersion of the $vp_{I}$ is shown in (b) and (c). The dispersions are obtained for $\kappa = 0.05$.}}
	\label{fig:vpI_hop}
\end{figure*}

\new{In an AA stacked model,} an {\em inter-layer vison pair}, $vp_I$, is made of two visons from adjacent layers, as shown in Fig.~\ref{fig:illustration}b. It is always confined to move in a plane (as the vison parity of a given plane cannot change) but can hop between nearest neighbour plaquettes within each layer due to the inter-layer perturbation $\Delta H_\perp$. The $vp_I$ pair thus moves on a  triangular lattice formed by the plaquettes of the honeycomb lattice.

We can now calculate the hopping matrix elements of an interlayer pair across an $\alpha$-bond, $t^{\alpha}_{vp_{I}},$ induced by 
$\Delta H^{AA}_\perp$ to leading order in $J_\perp$. For a vison pair in the layers $1$ and $2$ hopping across a $z$ bond we find
{\small
	\begin{align}
	\frac{t^{z}_{vp_{I}}}{J_\perp} = &\bra{\Phi^{2}_0(\vec R)}\otimes \bra{\Phi^1_0(\vec R)}\Delta H^z_\perp \ket{\Phi^{1}_0(\vec R')}\otimes \ket{\Phi^{2}_0(\vec R')}\\ \nonumber
	= &\bra{\Phi^1_0(\vec R)}\sigma^z_{i,1}\ket{\Phi^{1}_0(\vec R')}^2+\bra{\Phi^1_0(\vec R)}\sigma^z_{j,1}\ket{\Phi^{1}_0(\vec R')}^2,
	\label{eq:vp1_hop}
	\end{align}}
where $i$ and $j$ refer to the sites sharing a $z$-link.
The second equality is due to the decoupled nature of the layers in the unperturbed limit. The hopping amplitude is thus directly related to a magnetic field induced hopping rate of a vison in a monolayer Kitaev model. The single vison hopping rate was calculated in \citep{prx_vison,inti} and it was  found that in the gapless phase, for a single vison hopping due to a $\sigma^\alpha$ operator, a quasi-bound Majorana state leads to strong finite size effects. To avoid this issue, we open a small gap $\kappa$ in the Majorana spectrum by adding a time reversal symmetry breaking $nnn$ hopping. 
The calculation for one layer is done using periodic boundary conditions in a finite size system of width $L$. Thus, we are forced to consider two visons per layer at a distance of approximately $L/2$. For $\kappa \to 0$ 
the results depends sensitively on $L$, the precise location of the two visons and the direction of hopping as the quasi-bound Majorana states of the two vison pairs overlap. No such problem exist for sufficiently large $\kappa$ as shown in 
Fig.~\ref{fig:vpI_hop}a.  At a small $\kappa \approx 0.02 K$, the hopping amplitude can be read off from the plot as 
\begin{align}
t^{x}_{vp_I}=t^{y}_{vp_I}=t^{z}_{vp_I}\approx 0.35 ~J_\perp
\end{align}

Using this, we can calculate the dispersion of $vp_{I}$ as shown in Fig.~\ref{fig:vpI_hop}b. The excitation gap evolves as $\Delta_{vp_{I}}\approx E^0_{vp_{I}}-6~|t_{vp_{I}}|$ for $J<0$ with a band minimum at the $\Gamma$ point. For $J>0$, the gap is given by $\Delta_{vp_{I}}\approx E^0_{vp_{I}}-3~|t_{vp_{I}}|$ with multiple band minima at the $K$ points of the Brillouin zone.

It is interesting to point out that the dynamics of the inter-layer vison pair induced by the inter-layer Heisenberg coupling $J_\perp$ is strikingly different from the dynamics of a single vison induced by an external magnetic field. In both case the same type of $\sigma^z$ matrix elements $A_i=\bra{\Phi^1_0(\vec R)}\sigma^z_{i,1}\ket{\Phi^{1}_0(\vec R')}$ needs to be calculated. The single vison hopping is then determined by $A_i+A_j$ while the hopping rate of the pair arises from $A_i^2+A_j^2$.
For $K>0$ one has $A_i=-A_j$ and thus destructive interference prohibits field-induced single vison hopping 
\citep{prx_vison,inti}. This effect is absent for the inter-layer-pair hopping. For antiferromagnetic Kitaev coupling, $K>0$ and finite $\kappa$, the single-vison band is topological as has been shown in Ref.~\citep{prx_vison,inti}. Again this effect is absent for the inter-layer pair.

It should also be mentioned that there exists a second type of inter-layer vison pair, shown in Fig~\ref{fig:illustration}d which also acquires dynamics due to $\Delta H_\perp$. This excitation is bound to a bilayer and does not delocalize in space. However, it has an internal dynamics arising from the coupling (linear in $J_\perp$) between two states that differ by their vison positions as shown in Fig~\ref{fig:illustration}d. We do not discuss this particle further in this work.

\subsection{\new{Inter-layer vison pairs - AB stacking}}
\begin{figure*}[t]
	\centering
	\includegraphics[width=\linewidth]{ 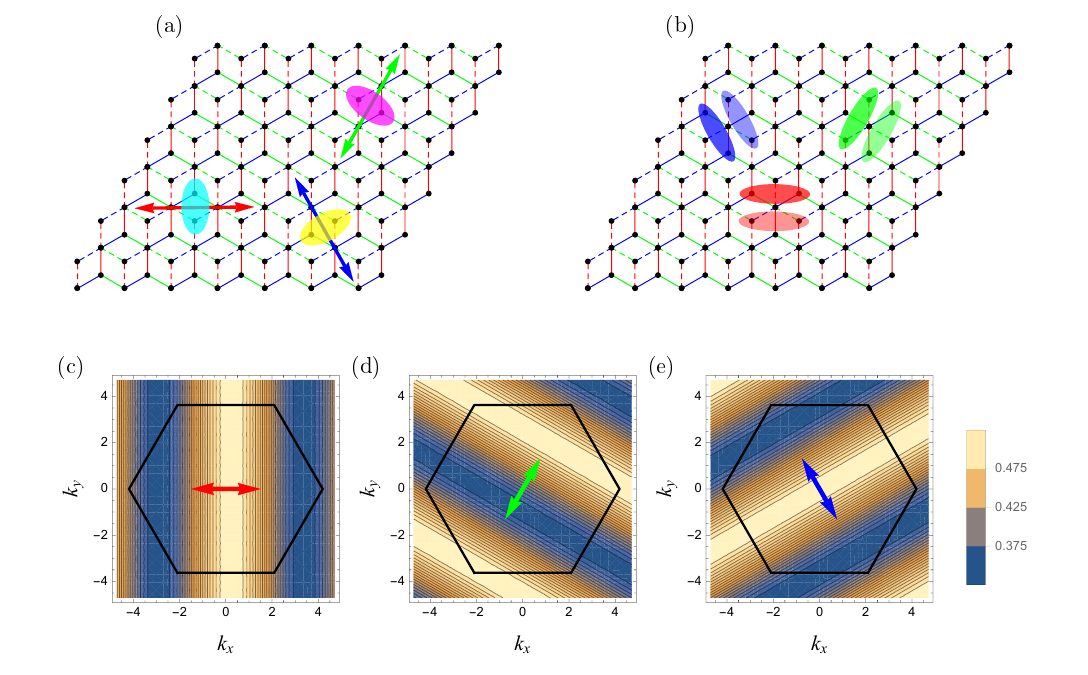}
	\caption{\small{\new{\textbf{Dynamics of vison pairs - AB stacking.} (a) Inter-layer vison pairs are illustrated for an AB stacked bilayer using ellipses that cover the centers of the two visons (one from each layer). Each of the three species, $vp^{AB}_{xy}$ (cyan), $vp^{AB}_{yz}$ (yellow) and $vp^{AB}_{zx}$ (violet)), is forced to move along the corresponding sheet operator (shown in Fig.~\ref{fig:ABloops}), indicated by the arrows. Figure (b) illustrates intra-layer vison pairs that can only move out of plane along the $z$ axis. The interlayer coupling induces hopping between the lightly shaded visons in the lower layer (dashed lines) and the darker shaded pair in the upper layer (solid lines), thus tracing a zig-zag path along the $z$ axis. Lower panel shows dispersions of the three inter-layer vison pairs, $vp^{AB}_{xy}$-(c), $vp^{AB}_{zx}$-(d) and $vp^{AB}_{yz}$-(e). The black solid line shows the first Brillouin zone. The one-dimensional nature of the excitations is highlighted by the colored arrows consistent with (a). The plots are obtained for $J_\perp=.1 $ and $\kappa=0.05$.}}}
	\label{fig:ABsublattice}
\end{figure*}
\new{The model with AA stacking discussed above differs qualitatively from the one with AB stacking. While for AA stacking, the vison stack operator, Eq.~\eqref{eq:visonStackOp}, are conservation laws, we found for AB stacking that instead sheet operators, Eq.~\eqref{eq:sheets}, are conserved. We will show that this difference has a profound influence on the dynamics of inter-layer vison pairs and their quantum numbers.
	
	Using the sheet conservation laws, one can identify three species of inter-layer pairs, as illustrated in Fig.~\ref{fig:ABsublattice}b. Remarkably, a given inter-layer pair is restricted to move along a one-dimensional channel within the 2D plane! The 1D channel is essentially the sheet operator of the bilayer and thus the three species $vp^{AB}_{xy}$, $vp^{AB}_{yz}$ and $vp^{AB}_{zx}$ propagate along the sheets $S^{xy}_m$, $S^{yz}_m$ and $S^{zx}_m$ respectively for a given index $m$. This sub-dimensional mobility is a direct consequence of the conservation laws (Eq.~\eqref{eq:sheets}) which prevent single visons from crossing a sheet.
	
	The hopping amplitudes are evaluated using the same methods as in the AA case leading to the relation
	\begin{align}
	t_{vp^{AB}_{\alpha\beta}} &= J_\perp\bra{\Phi^1_0(\vec R)}\sigma^\gamma_{i,1}\ket{\Phi^{1}_0(\vec R')}\bra{\Phi^2_0(\vec R)}\sigma^\gamma_{j,2}\ket{\Phi^{2}_0(\vec R')}
	\end{align}
	where $i \in $ sublattice A, $j \in$ sublattice B and $\alpha \ne \beta \ne \gamma$. The magnitude of the hopping rate is exactly half of that of the AA-model computed in Eq.~\eqref{eq:vp1_hop}, 
	\begin{align} |t_{vp^{AB}_{\alpha\beta}}|= \frac{|t^\gamma_{vp_I}|}{2}. \end{align}	The factor of $\frac{1}{2}$ relating the AA and AB stacked models simply arises from the fact that only one out of the two sublattices of a given monolayer unit cell contributes to the interlayer coupling in the AB model while both the sublattices additively contribute to the hopping in the AA case.
	
	The corresponding dispersions are plotted in Fig. \ref{fig:ABsublattice} showing their 1D nature which is to be contrasted with the fully 2D dispersion of inter-layer vison pairs in the AA model (Fig. \ref{fig:vpI_hop}c).
}

\begin{figure*}[t]
	\centering
	\includegraphics[width=\linewidth]{ 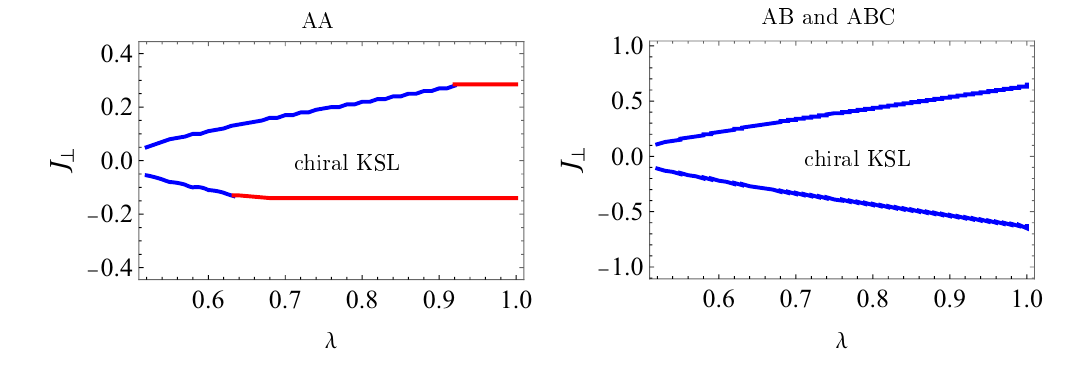}
	\caption{\small {
			Gap closing instabilities of the chiral KSL as a function of the anisotropy $\lambda$ and the inter-layer Heisenberg coupling $J_\perp$. \new{The left panel applies to AA stacking, the right panel to both AB and ABC stacking. } Vison-pair excitations become gapless when crossing the solid lines (estimated from perturbation theory linear in $J_\perp$). Across blue lines, the intra-layer pairs close the gap while the red lines indicate the gap closing due to inter-layer pairs ($vp_I$.). This plot is obtained for $\kappa = 0.05$. } }
	\label{fig:instability}
\end{figure*}
\subsection{\new{Intra-layer vison pair - AB stacking}}
\new{For the AB stacked model, similar to the AA case, intra-layer vison pairs $vp^{AB}_\alpha$ sharing an $\alpha$ bond in a layer can only hop between layers. In-plane motion of such pairs are forbidden by the sheet conservation laws discussed earlier. This hopping process is illustrated for a bilayer in Fig. \ref{fig:ABsublattice}b where three different species of intra-layer vison pairs that share $x$, $y$ and $z$ bonds are shown. As in the AA case, near the isotropic point $\lambda \apprge 0.72$, the excitations can only incoherently tunnel via Majorana assisted hopping with a diffusion constant
	\begin{align}
	D_{\perp}^{AB} = \frac{D_\perp }{2}
	\end{align}
	For $\lambda \apprle 0.72$, however, the intra-layer vison pair $vp^{AB}_z$ can coherently hop with a hopping amplitude
	\begin{align}
	t_{vp^{AB}_\alpha} =\frac{t_{vp_\alpha}}{2}
	\label{eq:vp2_hop_AB}
	\end{align}
	The factor of $\frac{1}{2}$ relating the AA and AB stacked models has already been discussed above. While the dispersions of the AA and AB intra-layer pairs are one-dimensional in the $z$ axis, the $vp^{AB}_z$ traces a zig-zag path in real space.
	
	For AA stacking we have discussed the dynamics of intra-layer vison pairs  (Ising anyon pairs) in the time-reversal broken phase. Here, all results apply directly also to the AB stacked model (up to the extra factor of $\frac{1}{2}$ in the hopping amplitudes).
}

\subsection{\new{Dynamics in ABC stacked model}}
\begin{figure*}[t]
	\centering
	\includegraphics[width=\linewidth]{ 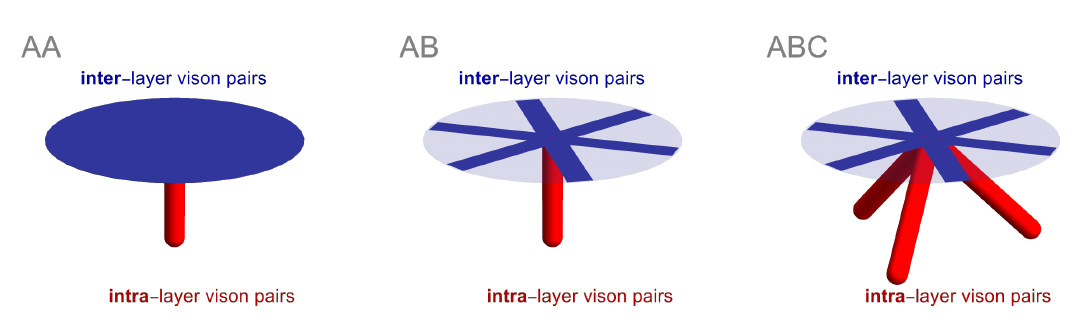}
	\caption{\new{Schematic illustration of the propagation of vison-pair excitations after a spot on the surface of a crystal is heated by a laser pulse.
			The laser pulse creates locally a high density of vison excitations. Vison pairs are mobile and move in a pattern characteristic of the dimensionality of their motion, see table~\ref{tab:summary}.} }
	\label{fig:response}
\end{figure*}

\new{The dynamics of inter-layer vison pairs in an ABC stacked model is identical to that of the AB model. This follows from the fact that a pair of neighboring layers in the ABC model is (up to translations) equivalent to a bilayer in the AB model.
 For an inter-layer vison pair in a given bilayer, the sheet conservation law restricts its motion to a 1D channel along the line obtained by projecting the sheet on to the bilayer. At linear order in $J_\perp$, the hopping is induced by the coupling between the two layers.
 \begin{align}
 t_{vp^{ABC}_{\alpha\beta}}=t_{vp^{AB}_{\alpha\beta}}
 \end{align}
An intra-layer pair, however, is affected qualitatively by the ABC stacking. In this case, the sheet operators are tilted. 
For a given intra-layer pair, there are two relevant sheet operators which cut the pair into half. 
Sheet-operator conservation then enforces that the pair moves parallel to these two sheets. Thus it can only move along the line obtained by the crossing of the two sheets, see Fig.~\ref{fig:ABloops}. 

Importantly, three types of intra-layer pairs ($vp^{ABC}_x$, $vp^{ABC}_y$ and $vp^{ABC}_y$) now move in three different directions, $(0,\sqrt{3}a,3 d_\perp ), (-\frac{\sqrt{3}}{2} a,-\frac{3}{2} a,3 d_\perp), (\frac{\sqrt{3}}{2} a,-\frac{3}{2} a,3 d_\perp)$, where $d_\perp$ is the distance of layers, see Fig.~\ref{fig:response}.  The hopping amplitude is however identical to that in the AB model due to the same arguments discussed above.
 \begin{align}
	t_{vp^{ABC}_{\alpha}}=t_{vp^{AB}_{\alpha}}
\end{align}}
\subsection{Instabilities}
Having computed the energy dispersions for the lowest energy gauge field excitations, we can estimate the critical $J_\perp$ at which the bands touch zero energy, thus destroying the Kitaev spin liquid phase. 

The resulting estimate for the location of the phase transition is shown in Fig.~\ref{fig:instability} where we assumed a small but finite Majorana gap, $\kappa=0.05$, to avoid large finite-size effects. The calculation proceeds in two steps. First, one determines, for $J_\perp=0$, the energy of the vison pair. Second, one obtains its hopping rate and thus the correction to the dispersion $E_k$ linear in $J_\perp$. Solving for $E_k=0$ at the band minimum allows to determine the critical $J_\perp$. 
As shown in Fig.~\ref{fig:instability}, the instability occurs at relatively small values of $J_\perp$ and thus likely in a parameter regime where the perturbative approach is approximately valid.

The color in Fig.~\ref{fig:instability} encodes the type of vison pair which drives the instability. 
For isotropic Kitaev couplings, $\lambda=1$, the leading instability \new{for both AB and ABC stacking  arises from the motion of intra-layer vison pairs while for AA stacking it is dominated by inter-layer pairs. This is because only for AA stacking do the inter-layer pairs have a two-dimensional dispersion and thus gain much more kinetic energy.} For large anisotropies  (smaller $\lambda$) in the AA stacked model, one can also find regimes, where the 1D motion of intra-layer pairs can trigger phase transitions.

A recent study by Vijayvargia, Seifert and Erten \citep{Urbanbilayer2} investigated, using mean-field theory, the coupling of an FM and an AFM Kitaev models by an Ising interaction.
They found that this model is more stable to inter-layer interactions compared to models where the two layers have the same sign for $K$. This is consistent with our results.
For the model of Ref.~\citep{Urbanbilayer2}, we actually obtain a vanishing inter-layer-pair hopping rate, $t_{vp_I}=0$, due to the same interference effect which was discussed above.
Thus, the inter-layer-pair hopping cannot easily drive a phase transition in this case.

\new{\subsection{Interplay with intra-layer perturbations}\label{sec:otherPert}
In this work, we considered pure Kitaev-layers coupled by Heisenberg interaction. While the inter-layer couplings can be sizeable \cite{Balz19}, see discussion below, in most realistic systems, we expect that other perturbations like Heisenberg coupling, Zeeman fields, or off-diagonal exchange interactions ($\Gamma$ term) {\em within} the layers are at least equally important. For single visons, we have investigated such models in Ref.~\cite{prx_vison}. For intra-layer vison pairs a comprehensive study was done by Zhang {\it et al.} \cite{Batista}.

Any extra perturbation will typically violate the stack- and sheet conservation laws of our models.
Consider, for example,  an inter-layer pair. 
We showed that it obtains a dispersion linear in the inter-layer Heisenberg coupling $J_\perp$. Intra-layer perturbations are able to move single visons as the relevant stack and sheet conservation laws are broken. Whether the single-vison hopping $t_v$ occurs to linear or quadratic order in the intra-layer coupling depends on the type of coupling and the sign of the Kitaev coupling as discussed in Ref.~\cite{prx_vison}.
The resulting hopping rate of the vison pair will be of order $t_v^2/E_{vp_I}$, where $E_{vp_I}$ is the vison-pair binding energy.

For intra-layer vison pairs, in contrast, the in-plane dynamics arises only from intra-layer perturbations, while $J_\perp$ is responsible for the motion across the layers.
}

\section{DISCUSSIONS}
In Kitaev materials the size of the inter-layer coupling can be remarkably large. In $\alpha-$RuCl$_3$, for example,  of the order of $1$\,meV \citep{Balz19,Jansen20} 
and thus of similar size as, e.g., ordering temperatures or Zeeman fields. Thus, it is important to investigate how inter-layer coupling may affect the properties and quantum numbers of excitations in Kitaev matter.

Here, we analysed two simplified three-dimensional models where Kitaev layers are coupled by a weak Heisenberg interaction.
In these models, the emergent $\mathbb Z_2$ gauge field  acquires dynamics. The lowest energy dynamical degrees of freedom are vison pairs since single visons are kept immobile by topology and conservation laws. Such bound vison-pairs can form within a single layer or between two adjacent layers and have qualitatively different dynamics and band dispersions. 

The intra-layer vison pairs, $vp_\alpha$, in the gapless isotropic Kitaev phase, for example, can move perpendicular to the layers only by incoherent Majorana-assisted hopping processes, as we showed from a combination of analytical and numerical arguments.

In contrast, an inter-layer vison pairs, $vp_I$ are confined to move within a plane only \new{and has a 2D dispersion when the layers are AA stacked. 
	For AB and ABC stacking, however, a novel sheet-conservation law, Eq.~\eqref{eq:sheets}, allows only for a one-dimensional motion of these excitations.
}

\new{When discussing the dimensionality of excitations, it is important to distinguish between the constraints that arise from topology (and are thus robust) and the ones arising from model-specific conservation laws, see also Sec. \ref{sec:otherPert}. Topology enforces the that inter-layer vison pairs can only move in a layer (as the vison-parity of each layer cannot be changed locally) while the sheet-conservation law makes their dynamics one-dimensional in our AB or ABC stacked model. This has to be contrasted with the physics of ``fractons'' where one-dimensional motion (or no motion at all) can arise in certain topological phases \cite{slagle17,fractonreview,fractonPretko}.}

As vison pairs can move coherently and gain a substantial amount of kinetic energy linear in $J_\perp$, they are prime candidates to trigger phase transitions when $J_\perp$ is increased, as discussed in the section of instabilities.

Inter-layer pairs have intriguing statistical exchange properties, best analyzed in the gapped chiral spin liquid phase. In this case, they carry two zero-energy Majorana modes, one per layer, giving rise to a rich set of fusion rules if two vison-pairs of the same bilayer meet. \new{Let us first consider AA stacking.} For a pair in layers $l$ and $l+1$ one finds four different fusion outcomes, 
$vp_I \times vp_I=\mathbb 1+ \psi_l + \psi_{l+1}+\psi_l \psi_{l+1}$, where $\psi_l$ denotes a fermionic excitation in layer $l$. A completely different set of fusion rules is obtained when a pair in layer $l-1$ and $l$ meets a pair in layer $l$ and $l+1$. Their fusion leads to single visons in layer $l-1$ and $l+2$, possibly releasing a fermion in layer $l$.

\new{For AB stacking, let us consider for simplicity only the case when an $xy$ inter-layer vison pair, $vp^{AB}_{xy,l}$, meets $vp^{AB}_{yz,l}$ in the same bilayer $l$. Due to the relative offset of the visons in the bilayer, only two of the four visons annihilate either in layer $l$ or $l+1$, leaving behind an intra-layer pair $vp^{AB}_y$, able to move in the $z$ direction and either one or zero Majorana modes,
	$vp^{AB}_{xy,l} \times vp^{AB}_{yz,l}=vp^{AB}_y (\mathbb 1+ \psi_l)+vp^{AB}_y (\mathbb 1+ \psi_{l+1})$. Identical results are obtained for ABC stacking, the only difference being that intra-layer pair $vp^{ABC}_y$  move in a  direction tilted relative to the $z$ axis.}

An interesting open question is the nature of the phases resulting from the proliferation of these excitations \cite{kyusungPRB}. In a phase, where inter-layer visons condense, one may obtain, for example, a situation where the single vison, is now able to hop from layer to layer, as the   
fusion of a single vison in layer $l$ with a inter-layer vison pair in layers $l$ and $l+1$ may result in a single vison in layer $l+1$.

\new{The coherent propagation of anyon pairs in the out-of-plane direction in the chiral spin liquid phase may be probed using the momentum dependent inelastic neutron spectroscopy. This should result in a sharp signature in the $(0,0,1)$ direction \cite{kohno2007spinons}. However, here one has to carefully separate the multi-particle continuum arising due to the presence of intra-layer perturbations which are of similar or larger strength compared to inter-layer couplings.
}
One way to investigate the peculiar mixture of the two-dimensional and three-dimensional dynamics in coupled Kitaev layers is to study what happens when one heats a spot on the surface of a Kitaev material locally by a laser. This will excite  Majorana modes, visons, and various bound states. Here, one can use the result that these modes have markedly different dynamics\new{, see Fig.~\ref{fig:response}.
Inter-layer pairs, for example, will diffuse parallel to the plane for AA stacking or along 1D channels within the plane for AB and ABC stacking. Intra-layer pairs move in perpendicular direction for AA and AB stacking but in a tripod-like pattern for ABC stacking as sketched in Fig.~\ref{fig:response}. 
Here, one also has to take into account other perturbations of the Kitaev model which trigger in-plane motion of single visons  \citep{prx_vison}.
Tracking the dynamics of such a system after a localized laser pulse, e.g., by local optical measurements
\citep{versteeg22}, will give insights into the rich and diverse dynamics of excitations in Kitaev materials.}

\section{METHODS}
\subsection{Matrix elements for perturbation theory}
In order to compute the hopping matrix elements of vison-pairs in Eq.~\eqref{eq:matrix_element}, we need the exact eigenstates in the unperturbed system in the limit $J_\perp \to 0$.
For the analysis of different vison-pair states, we need the ground-state wavefunction of layer $l$ assuming that $n$ visons  are located at positions $\vec R_1,\dots, \vec R_n$ within layer $l$.
The wavefunction can be expressed as a direct product of the matter fermion ($c_i$) and gauge fermion sector ($b_i$). 
\begin{align}
\ket{\Phi^{l}_0(\{\vec R\})}= \hat P_l \ket{\mathcal G(\{\vec R_l\})}_l\ket{M(\mathcal G)}_l
\end{align}
with $\{\vec R\}=\{\vec R_1,\dots, \vec R_n\}$ while $\ket{\mathcal G(\{\vec R_l\})}$ describes the wave function of the gauge fermions. $\ket{M(\mathcal G)}_l$ is the many-body wavefunction of the matter fermions in the gauge $\mathcal G$ determined by the $b$ fermion wavefunction. The operator $\hat P$ symmetrizes over all gauge transformations and projects the states onto the physical Hilbert space.
Here it is important to realize that for a given flux configuration only half of the matter-fermion states are physical, i.e., survive projection. Whether the parity of matter fermions is even or odd is a gauge-dependent statement. For practical calculations, one can, for example, follow Ref.~\citep{Loss,vojta} and calculate the total parity (matter+gauge fermions) to identify physical states. 

The exact wave functions of the decoupled multilayer system ($N$ layers) can then be written as a tensor product\\ $\ket{\Psi(\{\vec R_1\},\{\vec R_2\},..,\{\vec R_N\})}=\otimes_l\ket{\Phi^l(\{\vec R_l\})}$.

Since Eq.~\eqref{eq:vp2_hop} and Eq.~\eqref{eq:vp1_hop} are expressed in terms of single layer wavefunctions, we only need to solve a single layer Hamiltonian. In our calculations we use finite systems (monolayer) of sizes up to $N =2L^2 \sim 20000$ to obtain the many-body wavefunctions through diagonalization of the Majorana fermionic Hamiltonian (Eq.~\eqref{eq:majorana_Ham}). We use periodic boundary conditions to avoid spurious edge modes which then forces us to add visons only in pairs (vison number is conserved modulo 2). For the inter-layer pair wavefunctions, Eq.~\eqref{eq:vp1_hop}, we therefore add two visons which are separated by approximately $L/2$ in a system of linear size $L$. The matrix element is calculated by moving only one of the visons, keeping the second one fixed. The wavefunction overlaps are then computed using the Pfaffian determinant method described in Refs.~\citep{robledo,prx_vison}.

\section*{DATA AVAILABILITY}
The datasets generated and/or analysed in this study are available from Zenodo \cite{dataZenodo}.
\section*{CODE AVAILABILITY}
The codes used during the current study will be available from the corresponding author at reasonable request.
\section*{ACKNOWLEDGEMENTS}
We would like to thank Urban Seifert for useful discussions.
This work was supported by the
Deutsche Forschungsgemeinschaft (DFG) through CRC1238 (Project
No. 277146847, project C02 and C04) and by the Bonn-Cologne Graduate School of Physics and Astronomy (BCGS).

\section*{AUTHOR CONTRIBUTIONS}
APJ and AR designed the study. APJ carried out the calculations with input from AR. APJ and AR wrote the manuscript together.
\section*{CONFLICTING INTERESTS}
The authors declare no competing interests.
\bibliography{vison_multilayer}

\providecommand{\noopsort}[1]{}\providecommand{\singleletter}[1]{#1}%
\begin{thebibliography}{46}%
\makeatletter
\providecommand \@ifxundefined [1]{%
 \@ifx{#1\undefined}
}%
\providecommand \@ifnum [1]{%
 \ifnum #1\expandafter \@firstoftwo
 \else \expandafter \@secondoftwo
 \fi
}%
\providecommand \@ifx [1]{%
 \ifx #1\expandafter \@firstoftwo
 \else \expandafter \@secondoftwo
 \fi
}%
\providecommand \natexlab [1]{#1}%
\providecommand \enquote  [1]{``#1''}%
\providecommand \bibnamefont  [1]{#1}%
\providecommand \bibfnamefont [1]{#1}%
\providecommand \citenamefont [1]{#1}%
\providecommand \href@noop [0]{\@secondoftwo}%
\providecommand \href [0]{\begingroup \@sanitize@url \@href}%
\providecommand \@href[1]{\@@startlink{#1}\@@href}%
\providecommand \@@href[1]{\endgroup#1\@@endlink}%
\providecommand \@sanitize@url [0]{\catcode `\\12\catcode `\$12\catcode
  `\&12\catcode `\#12\catcode `\^12\catcode `\_12\catcode `\%12\relax}%
\providecommand \@@startlink[1]{}%
\providecommand \@@endlink[0]{}%
\providecommand \url  [0]{\begingroup\@sanitize@url \@url }%
\providecommand \@url [1]{\endgroup\@href {#1}{\urlprefix }}%
\providecommand \urlprefix  [0]{URL }%
\providecommand \Eprint [0]{\href }%
\providecommand \doibase [0]{https://doi.org/}%
\providecommand \selectlanguage [0]{\@gobble}%
\providecommand \bibinfo  [0]{\@secondoftwo}%
\providecommand \bibfield  [0]{\@secondoftwo}%
\providecommand \translation [1]{[#1]}%
\providecommand \BibitemOpen [0]{}%
\providecommand \bibitemStop [0]{}%
\providecommand \bibitemNoStop [0]{.\EOS\space}%
\providecommand \EOS [0]{\spacefactor3000\relax}%
\providecommand \BibitemShut  [1]{\csname bibitem#1\endcsname}%
\let\auto@bib@innerbib\@empty
%</preamble>
\bibitem [{\citenamefont {Senthil}\ and\ \citenamefont
  {Fisher}(2000)}]{visonSenthil}%
  \BibitemOpen
  \bibfield  {author} {\bibinfo {author} {\bibfnamefont {T.}~\bibnamefont
  {Senthil}}\ and\ \bibinfo {author} {\bibfnamefont {M.~P.~A.}\ \bibnamefont
  {Fisher}},\ }\bibfield  {title} {\bibinfo {title} {{${Z}_{2}$ gauge theory of
  electron fractionalization in strongly correlated systems}},\ }\href
  {https://doi.org/10.1103/PhysRevB.62.7850} {\bibfield  {journal} {\bibinfo
  {journal} {Phys. Rev. B}\ }\textbf {\bibinfo {volume} {62}},\ \bibinfo
  {pages} {7850} (\bibinfo {year} {2000})}\BibitemShut {NoStop}%
\bibitem [{\citenamefont {Zhou}\ \emph {et~al.}(2017)\citenamefont {Zhou},
  \citenamefont {Kanoda},\ and\ \citenamefont {Ng}}]{ZhouQSL}%
  \BibitemOpen
  \bibfield  {author} {\bibinfo {author} {\bibfnamefont {Y.}~\bibnamefont
  {Zhou}}, \bibinfo {author} {\bibfnamefont {K.}~\bibnamefont {Kanoda}},\ and\
  \bibinfo {author} {\bibfnamefont {T.-K.}\ \bibnamefont {Ng}},\ }\bibfield
  {title} {\bibinfo {title} {Quantum spin liquid states},\ }\href
  {https://doi.org/10.1103/RevModPhys.89.025003} {\bibfield  {journal}
  {\bibinfo  {journal} {Rev. Mod. Phys.}\ }\textbf {\bibinfo {volume} {89}},\
  \bibinfo {pages} {025003} (\bibinfo {year} {2017})}\BibitemShut {NoStop}%
\bibitem [{\citenamefont {Savary}\ and\ \citenamefont
  {Balents}(2016)}]{savaryBalents}%
  \BibitemOpen
  \bibfield  {author} {\bibinfo {author} {\bibfnamefont {L.}~\bibnamefont
  {Savary}}\ and\ \bibinfo {author} {\bibfnamefont {L.}~\bibnamefont
  {Balents}},\ }\bibfield  {title} {\bibinfo {title} {Quantum spin liquids: a
  review},\ }\href@noop {} {\bibfield  {journal} {\bibinfo  {journal} {Reports
  on Progress in Physics}\ }\textbf {\bibinfo {volume} {80}},\ \bibinfo {pages}
  {016502} (\bibinfo {year} {2016})}\BibitemShut {NoStop}%
\bibitem [{\citenamefont {Kitaev}(2006)}]{Kitaev06}%
  \BibitemOpen
  \bibfield  {author} {\bibinfo {author} {\bibfnamefont {A.}~\bibnamefont
  {Kitaev}},\ }\bibfield  {title} {\bibinfo {title} {{Anyons in an exactly
  solved model and beyond}},\ }\href
  {https://doi.org/https://doi.org/10.1016/j.aop.2005.10.005} {\bibfield
  {journal} {\bibinfo  {journal} {Annals of Physics}\ }\textbf {\bibinfo
  {volume} {321}},\ \bibinfo {pages} {2 } (\bibinfo {year} {2006})}\BibitemShut
  {NoStop}%
\bibitem [{\citenamefont {Jackeli}\ and\ \citenamefont
  {Khaliullin}(2009)}]{JackeliKhaliullin}%
  \BibitemOpen
  \bibfield  {author} {\bibinfo {author} {\bibfnamefont {G.}~\bibnamefont
  {Jackeli}}\ and\ \bibinfo {author} {\bibfnamefont {G.}~\bibnamefont
  {Khaliullin}},\ }\bibfield  {title} {\bibinfo {title} {Mott insulators in the
  strong spin-orbit coupling limit: From heisenberg to a quantum compass and
  kitaev models},\ }\href {https://doi.org/10.1103/PhysRevLett.102.017205}
  {\bibfield  {journal} {\bibinfo  {journal} {Phys. Rev. Lett.}\ }\textbf
  {\bibinfo {volume} {102}},\ \bibinfo {pages} {017205} (\bibinfo {year}
  {2009})}\BibitemShut {NoStop}%
\bibitem [{\citenamefont {Banerjee}\ \emph {et~al.}(2018)\citenamefont
  {Banerjee}, \citenamefont {Lampen-Kelley}, \citenamefont {Knolle},
  \citenamefont {Balz}, \citenamefont {Aczel}, \citenamefont {Winn},
  \citenamefont {Liu}, \citenamefont {Pajerowski}, \citenamefont {Yan},
  \citenamefont {Bridges}, \citenamefont {Savici}, \citenamefont {Chakoumakos},
  \citenamefont {Lumsden}, \citenamefont {Tennant}, \citenamefont {Moessner},
  \citenamefont {Mandrus},\ and\ \citenamefont {Nagler}}]{fieldinduced}%
  \BibitemOpen
  \bibfield  {author} {\bibinfo {author} {\bibfnamefont {A.}~\bibnamefont
  {Banerjee}}, \bibinfo {author} {\bibfnamefont {P.}~\bibnamefont
  {Lampen-Kelley}}, \bibinfo {author} {\bibfnamefont {J.}~\bibnamefont
  {Knolle}}, \bibinfo {author} {\bibfnamefont {C.}~\bibnamefont {Balz}},
  \bibinfo {author} {\bibfnamefont {A.~A.}\ \bibnamefont {Aczel}}, \bibinfo
  {author} {\bibfnamefont {B.}~\bibnamefont {Winn}}, \bibinfo {author}
  {\bibfnamefont {Y.}~\bibnamefont {Liu}}, \bibinfo {author} {\bibfnamefont
  {D.}~\bibnamefont {Pajerowski}}, \bibinfo {author} {\bibfnamefont
  {J.}~\bibnamefont {Yan}}, \bibinfo {author} {\bibfnamefont {C.~A.}\
  \bibnamefont {Bridges}}, \bibinfo {author} {\bibfnamefont {A.~T.}\
  \bibnamefont {Savici}}, \bibinfo {author} {\bibfnamefont {B.~C.}\
  \bibnamefont {Chakoumakos}}, \bibinfo {author} {\bibfnamefont {M.~D.}\
  \bibnamefont {Lumsden}}, \bibinfo {author} {\bibfnamefont {D.~A.}\
  \bibnamefont {Tennant}}, \bibinfo {author} {\bibfnamefont {R.}~\bibnamefont
  {Moessner}}, \bibinfo {author} {\bibfnamefont {D.~G.}\ \bibnamefont
  {Mandrus}},\ and\ \bibinfo {author} {\bibfnamefont {S.~E.}\ \bibnamefont
  {Nagler}},\ }\bibfield  {title} {\bibinfo {title} {{Excitations in the
  field-induced quantum spin liquid state of $\alpha$-RuCl$_3$}},\ }\href
  {https://doi.org/10.1038/s41535-018-0079-2} {\bibfield  {journal} {\bibinfo
  {journal} {npj Quantum Materials}\ }\textbf {\bibinfo {volume} {3}},\
  \bibinfo {pages} {8} (\bibinfo {year} {2018})}\BibitemShut {NoStop}%
\bibitem [{\citenamefont {Moretti~Sala}\ \emph {et~al.}(2020)\citenamefont
  {Moretti~Sala}, \citenamefont {Monaco}, \citenamefont {Hickey}, \citenamefont
  {Becker}, \citenamefont {Freund}, \citenamefont {Jesche}, \citenamefont
  {Gegenwart}, \citenamefont {Eschmann}, \citenamefont {Buessen}, \citenamefont
  {Trebst}, \citenamefont {van Loosdrecht}, \citenamefont {van~den Brink},
  \citenamefont {Grüninger},\ and\ \citenamefont {Revelli}}]{nairo3}%
  \BibitemOpen
  \bibfield  {author} {\bibinfo {author} {\bibfnamefont {M.}~\bibnamefont
  {Moretti~Sala}}, \bibinfo {author} {\bibfnamefont {G.}~\bibnamefont
  {Monaco}}, \bibinfo {author} {\bibfnamefont {C.}~\bibnamefont {Hickey}},
  \bibinfo {author} {\bibfnamefont {P.}~\bibnamefont {Becker}}, \bibinfo
  {author} {\bibfnamefont {F.}~\bibnamefont {Freund}}, \bibinfo {author}
  {\bibfnamefont {A.}~\bibnamefont {Jesche}}, \bibinfo {author} {\bibfnamefont
  {P.}~\bibnamefont {Gegenwart}}, \bibinfo {author} {\bibfnamefont
  {T.}~\bibnamefont {Eschmann}}, \bibinfo {author} {\bibfnamefont {F.~L.}\
  \bibnamefont {Buessen}}, \bibinfo {author} {\bibfnamefont {S.}~\bibnamefont
  {Trebst}}, \bibinfo {author} {\bibfnamefont {P.~H.~M.}\ \bibnamefont {van
  Loosdrecht}}, \bibinfo {author} {\bibfnamefont {J.}~\bibnamefont {van~den
  Brink}}, \bibinfo {author} {\bibfnamefont {M.}~\bibnamefont {Grüninger}},\
  and\ \bibinfo {author} {\bibfnamefont {A.}~\bibnamefont {Revelli}},\
  }\bibfield  {title} {\bibinfo {title} {{Fingerprints of Kitaev physics in the
  magnetic excitations of honeycomb iridates}},\ }\href
  {https://link.aps.org/doi/10.1103/PhysRevResearch.2.043094} {\bibfield
  {journal} {\bibinfo  {journal} {Physical Review Research}\ }\textbf {\bibinfo
  {volume} {2}} (\bibinfo {year} {2020})}\BibitemShut {NoStop}%
\bibitem [{\citenamefont {Kasahara}\ \emph {et~al.}(2018)\citenamefont
  {Kasahara}, \citenamefont {Ohnishi}, \citenamefont {Mizukami}, \citenamefont
  {Tanaka}, \citenamefont {Ma}, \citenamefont {Sugii}, \citenamefont {Kurita},
  \citenamefont {Tanaka}, \citenamefont {Nasu}, \citenamefont {Motome},
  \citenamefont {Shibauchi},\ and\ \citenamefont {Matsuda}}]{kasahara1}%
  \BibitemOpen
  \bibfield  {author} {\bibinfo {author} {\bibfnamefont {Y.}~\bibnamefont
  {Kasahara}}, \bibinfo {author} {\bibfnamefont {T.}~\bibnamefont {Ohnishi}},
  \bibinfo {author} {\bibfnamefont {Y.}~\bibnamefont {Mizukami}}, \bibinfo
  {author} {\bibfnamefont {O.}~\bibnamefont {Tanaka}}, \bibinfo {author}
  {\bibfnamefont {S.}~\bibnamefont {Ma}}, \bibinfo {author} {\bibfnamefont
  {K.}~\bibnamefont {Sugii}}, \bibinfo {author} {\bibfnamefont
  {N.}~\bibnamefont {Kurita}}, \bibinfo {author} {\bibfnamefont
  {H.}~\bibnamefont {Tanaka}}, \bibinfo {author} {\bibfnamefont
  {J.}~\bibnamefont {Nasu}}, \bibinfo {author} {\bibfnamefont {Y.}~\bibnamefont
  {Motome}}, \bibinfo {author} {\bibfnamefont {T.}~\bibnamefont {Shibauchi}},\
  and\ \bibinfo {author} {\bibfnamefont {Y.}~\bibnamefont {Matsuda}},\
  }\bibfield  {title} {\bibinfo {title} {{"Majorana quantization and
  half-integer thermal quantum Hall effect in a Kitaev spin liquid"}},\
  }\href@noop {} {\bibfield  {journal} {\bibinfo  {journal} {Nature}\ }\textbf
  {\bibinfo {volume} {559}} (\bibinfo {year} {2018})}\BibitemShut {NoStop}%
\bibitem [{\citenamefont {Yokoi}\ \emph {et~al.}(2021)\citenamefont {Yokoi},
  \citenamefont {Ma}, \citenamefont {Kasahara}, \citenamefont {Kasahara},
  \citenamefont {Shibauchi}, \citenamefont {Kurita}, \citenamefont {Tanaka},
  \citenamefont {Nasu}, \citenamefont {Motome}, \citenamefont {Hickey},
  \citenamefont {Trebst},\ and\ \citenamefont {Matsuda}}]{kasahara2}%
  \BibitemOpen
  \bibfield  {author} {\bibinfo {author} {\bibfnamefont {T.}~\bibnamefont
  {Yokoi}}, \bibinfo {author} {\bibfnamefont {S.}~\bibnamefont {Ma}}, \bibinfo
  {author} {\bibfnamefont {Y.}~\bibnamefont {Kasahara}}, \bibinfo {author}
  {\bibfnamefont {S.}~\bibnamefont {Kasahara}}, \bibinfo {author}
  {\bibfnamefont {T.}~\bibnamefont {Shibauchi}}, \bibinfo {author}
  {\bibfnamefont {N.}~\bibnamefont {Kurita}}, \bibinfo {author} {\bibfnamefont
  {H.}~\bibnamefont {Tanaka}}, \bibinfo {author} {\bibfnamefont
  {J.}~\bibnamefont {Nasu}}, \bibinfo {author} {\bibfnamefont {Y.}~\bibnamefont
  {Motome}}, \bibinfo {author} {\bibfnamefont {C.}~\bibnamefont {Hickey}},
  \bibinfo {author} {\bibfnamefont {S.}~\bibnamefont {Trebst}},\ and\ \bibinfo
  {author} {\bibfnamefont {Y.}~\bibnamefont {Matsuda}},\ }\bibfield  {title}
  {\bibinfo {title} {{Half-integer quantized anomalous thermal Hall effect in
  the Kitaev material candidate $\alpha$-RuCl$_3$}},\ }\href
  {http://science.sciencemag.org/content/373/6554/568.abstract} {\bibfield
  {journal} {\bibinfo  {journal} {Science}\ }\textbf {\bibinfo {volume} {373}}
  (\bibinfo {year} {2021})}\BibitemShut {NoStop}%
\bibitem [{\citenamefont {Vinkler-Aviv}\ and\ \citenamefont
  {Rosch}(2018)}]{RoschQuantumHall}%
  \BibitemOpen
  \bibfield  {author} {\bibinfo {author} {\bibfnamefont {Y.}~\bibnamefont
  {Vinkler-Aviv}}\ and\ \bibinfo {author} {\bibfnamefont {A.}~\bibnamefont
  {Rosch}},\ }\bibfield  {title} {\bibinfo {title} {{Approximately Quantized
  Thermal Hall Effect of Chiral Liquids Coupled to Phonons}},\ }\href
  {https://doi.org/10.1103/PhysRevX.8.031032} {\bibfield  {journal} {\bibinfo
  {journal} {Phys. Rev. X}\ }\textbf {\bibinfo {volume} {8}},\ \bibinfo {pages}
  {031032} (\bibinfo {year} {2018})}\BibitemShut {NoStop}%
\bibitem [{\citenamefont {Ye}\ \emph {et~al.}(2018)\citenamefont {Ye},
  \citenamefont {Hal\'asz}, \citenamefont {Savary},\ and\ \citenamefont
  {Balents}}]{Balents}%
  \BibitemOpen
  \bibfield  {author} {\bibinfo {author} {\bibfnamefont {M.}~\bibnamefont
  {Ye}}, \bibinfo {author} {\bibfnamefont {G.~B.}\ \bibnamefont {Hal\'asz}},
  \bibinfo {author} {\bibfnamefont {L.}~\bibnamefont {Savary}},\ and\ \bibinfo
  {author} {\bibfnamefont {L.}~\bibnamefont {Balents}},\ }\bibfield  {title}
  {\bibinfo {title} {{Quantization of the Thermal Hall Conductivity at Small
  Hall Angles}},\ }\href {https://doi.org/10.1103/PhysRevLett.121.147201}
  {\bibfield  {journal} {\bibinfo  {journal} {Phys. Rev. Lett.}\ }\textbf
  {\bibinfo {volume} {121}},\ \bibinfo {pages} {147201} (\bibinfo {year}
  {2018})}\BibitemShut {NoStop}%
\bibitem [{\citenamefont {Yamashita}\ \emph {et~al.}(2020)\citenamefont
  {Yamashita}, \citenamefont {Gouchi}, \citenamefont {Uwatoko}, \citenamefont
  {Kurita},\ and\ \citenamefont {Tanaka}}]{PhysRevB.102.220404}%
  \BibitemOpen
  \bibfield  {author} {\bibinfo {author} {\bibfnamefont {M.}~\bibnamefont
  {Yamashita}}, \bibinfo {author} {\bibfnamefont {J.}~\bibnamefont {Gouchi}},
  \bibinfo {author} {\bibfnamefont {Y.}~\bibnamefont {Uwatoko}}, \bibinfo
  {author} {\bibfnamefont {N.}~\bibnamefont {Kurita}},\ and\ \bibinfo {author}
  {\bibfnamefont {H.}~\bibnamefont {Tanaka}},\ }\bibfield  {title} {\bibinfo
  {title} {Sample dependence of half-integer quantized thermal hall effect in
  the kitaev spin-liquid candidate
  $\ensuremath{\alpha}\text{\ensuremath{-}}{\mathrm{rucl}}_{3}$},\ }\href
  {https://doi.org/10.1103/PhysRevB.102.220404} {\bibfield  {journal} {\bibinfo
   {journal} {Phys. Rev. B}\ }\textbf {\bibinfo {volume} {102}},\ \bibinfo
  {pages} {220404} (\bibinfo {year} {2020})}\BibitemShut {NoStop}%
\bibitem [{\citenamefont {Song}\ \emph {et~al.}(2016)\citenamefont {Song},
  \citenamefont {You},\ and\ \citenamefont {Balents}}]{Songprl}%
  \BibitemOpen
  \bibfield  {author} {\bibinfo {author} {\bibfnamefont {X.-Y.}\ \bibnamefont
  {Song}}, \bibinfo {author} {\bibfnamefont {Y.-Z.}\ \bibnamefont {You}},\ and\
  \bibinfo {author} {\bibfnamefont {L.}~\bibnamefont {Balents}},\ }\bibfield
  {title} {\bibinfo {title} {Low-energy spin dynamics of the honeycomb spin
  liquid beyond the kitaev limit},\ }\href
  {https://doi.org/10.1103/PhysRevLett.117.037209} {\bibfield  {journal}
  {\bibinfo  {journal} {Phys. Rev. Lett.}\ }\textbf {\bibinfo {volume} {117}},\
  \bibinfo {pages} {037209} (\bibinfo {year} {2016})}\BibitemShut {NoStop}%
\bibitem [{\citenamefont {Knolle}\ \emph {et~al.}(2018)\citenamefont {Knolle},
  \citenamefont {Bhattacharjee},\ and\ \citenamefont
  {Moessner}}]{KnolleAugmented18}%
  \BibitemOpen
  \bibfield  {author} {\bibinfo {author} {\bibfnamefont {J.}~\bibnamefont
  {Knolle}}, \bibinfo {author} {\bibfnamefont {S.}~\bibnamefont
  {Bhattacharjee}},\ and\ \bibinfo {author} {\bibfnamefont {R.}~\bibnamefont
  {Moessner}},\ }\bibfield  {title} {\bibinfo {title} {Dynamics of a quantum
  spin liquid beyond integrability: The
  kitaev-heisenberg-$\mathrm{\ensuremath{\Gamma}}$ model in an augmented parton
  mean-field theory},\ }\href {https://doi.org/10.1103/PhysRevB.97.134432}
  {\bibfield  {journal} {\bibinfo  {journal} {Phys. Rev. B}\ }\textbf {\bibinfo
  {volume} {97}},\ \bibinfo {pages} {134432} (\bibinfo {year}
  {2018})}\BibitemShut {NoStop}%
\bibitem [{\citenamefont {Lunkin}\ \emph {et~al.}(2019)\citenamefont {Lunkin},
  \citenamefont {Tikhonov},\ and\ \citenamefont
  {Feigel'man}}]{lunkin2019perturbed}%
  \BibitemOpen
  \bibfield  {author} {\bibinfo {author} {\bibfnamefont {A.}~\bibnamefont
  {Lunkin}}, \bibinfo {author} {\bibfnamefont {K.}~\bibnamefont {Tikhonov}},\
  and\ \bibinfo {author} {\bibfnamefont {M.}~\bibnamefont {Feigel'man}},\
  }\bibfield  {title} {\bibinfo {title} {Perturbed kitaev model: Excitation
  spectrum and long-ranged spin correlations},\ }\href@noop {} {\bibfield
  {journal} {\bibinfo  {journal} {Journal of Physics and Chemistry of Solids}\
  }\textbf {\bibinfo {volume} {128}},\ \bibinfo {pages} {130} (\bibinfo {year}
  {2019})}\BibitemShut {NoStop}%
\bibitem [{\citenamefont {Tikhonov}\ \emph {et~al.}(2011)\citenamefont
  {Tikhonov}, \citenamefont {Feigel'man},\ and\ \citenamefont
  {Kitaev}}]{Tikhonovprl}%
  \BibitemOpen
  \bibfield  {author} {\bibinfo {author} {\bibfnamefont {K.~S.}\ \bibnamefont
  {Tikhonov}}, \bibinfo {author} {\bibfnamefont {M.~V.}\ \bibnamefont
  {Feigel'man}},\ and\ \bibinfo {author} {\bibfnamefont {A.~Y.}\ \bibnamefont
  {Kitaev}},\ }\bibfield  {title} {\bibinfo {title} {Power-law spin
  correlations in a perturbed spin model on a honeycomb lattice},\ }\href
  {https://doi.org/10.1103/PhysRevLett.106.067203} {\bibfield  {journal}
  {\bibinfo  {journal} {Phys. Rev. Lett.}\ }\textbf {\bibinfo {volume} {106}},\
  \bibinfo {pages} {067203} (\bibinfo {year} {2011})}\BibitemShut {NoStop}%
\bibitem [{\citenamefont {Mandal}\ \emph {et~al.}(2011)\citenamefont {Mandal},
  \citenamefont {Bhattacharjee}, \citenamefont {Sengupta}, \citenamefont
  {Shankar},\ and\ \citenamefont {Baskaran}}]{mandal11}%
  \BibitemOpen
  \bibfield  {author} {\bibinfo {author} {\bibfnamefont {S.}~\bibnamefont
  {Mandal}}, \bibinfo {author} {\bibfnamefont {S.}~\bibnamefont
  {Bhattacharjee}}, \bibinfo {author} {\bibfnamefont {K.}~\bibnamefont
  {Sengupta}}, \bibinfo {author} {\bibfnamefont {R.}~\bibnamefont {Shankar}},\
  and\ \bibinfo {author} {\bibfnamefont {G.}~\bibnamefont {Baskaran}},\
  }\bibfield  {title} {\bibinfo {title} {Confinement-deconfinement transition
  and spin correlations in a generalized kitaev model},\ }\href
  {https://doi.org/10.1103/PhysRevB.84.155121} {\bibfield  {journal} {\bibinfo
  {journal} {Phys. Rev. B}\ }\textbf {\bibinfo {volume} {84}},\ \bibinfo
  {pages} {155121} (\bibinfo {year} {2011})}\BibitemShut {NoStop}%
\bibitem [{\citenamefont {Chen}\ and\ \citenamefont {Villadiego}(2023)}]{inti}%
  \BibitemOpen
  \bibfield  {author} {\bibinfo {author} {\bibfnamefont {C.}~\bibnamefont
  {Chen}}\ and\ \bibinfo {author} {\bibfnamefont {I.~S.}\ \bibnamefont
  {Villadiego}},\ }\bibfield  {title} {\bibinfo {title} {Nature of visons in
  the perturbed ferromagnetic and antiferromagnetic kitaev honeycomb models},\
  }\href {https://doi.org/10.1103/PhysRevB.107.045114} {\bibfield  {journal}
  {\bibinfo  {journal} {Phys. Rev. B}\ }\textbf {\bibinfo {volume} {107}},\
  \bibinfo {pages} {045114} (\bibinfo {year} {2023})}\BibitemShut {NoStop}%
\bibitem [{\citenamefont {Joy}\ and\ \citenamefont {Rosch}(2022)}]{prx_vison}%
  \BibitemOpen
  \bibfield  {author} {\bibinfo {author} {\bibfnamefont {A.~P.}\ \bibnamefont
  {Joy}}\ and\ \bibinfo {author} {\bibfnamefont {A.}~\bibnamefont {Rosch}},\
  }\bibfield  {title} {\bibinfo {title} {Dynamics of visons and thermal hall
  effect in perturbed kitaev models},\ }\href
  {https://doi.org/10.1103/PhysRevX.12.041004} {\bibfield  {journal} {\bibinfo
  {journal} {Phys. Rev. X}\ }\textbf {\bibinfo {volume} {12}},\ \bibinfo
  {pages} {041004} (\bibinfo {year} {2022})}\BibitemShut {NoStop}%
\bibitem [{\citenamefont {Zhang}\ \emph {et~al.}(2021)\citenamefont {Zhang},
  \citenamefont {Hal\'asz}, \citenamefont {Zhu},\ and\ \citenamefont
  {Batista}}]{Batista}%
  \BibitemOpen
  \bibfield  {author} {\bibinfo {author} {\bibfnamefont {S.-S.}\ \bibnamefont
  {Zhang}}, \bibinfo {author} {\bibfnamefont {G.~B.}\ \bibnamefont {Hal\'asz}},
  \bibinfo {author} {\bibfnamefont {W.}~\bibnamefont {Zhu}},\ and\ \bibinfo
  {author} {\bibfnamefont {C.~D.}\ \bibnamefont {Batista}},\ }\bibfield
  {title} {\bibinfo {title} {{Variational study of the Kitaev-Heisenberg-Gamma
  model}},\ }\href {https://doi.org/10.1103/PhysRevB.104.014411} {\bibfield
  {journal} {\bibinfo  {journal} {Phys. Rev. B}\ }\textbf {\bibinfo {volume}
  {104}},\ \bibinfo {pages} {014411} (\bibinfo {year} {2021})}\BibitemShut
  {NoStop}%
\bibitem [{\citenamefont {Balz}\ \emph {et~al.}(2021)\citenamefont {Balz},
  \citenamefont {Janssen}, \citenamefont {Lampen-Kelley}, \citenamefont
  {Banerjee}, \citenamefont {Liu}, \citenamefont {Yan}, \citenamefont
  {Mandrus}, \citenamefont {Vojta},\ and\ \citenamefont
  {Nagler}}]{PhysRevB.103.174417}%
  \BibitemOpen
  \bibfield  {author} {\bibinfo {author} {\bibfnamefont {C.}~\bibnamefont
  {Balz}}, \bibinfo {author} {\bibfnamefont {L.}~\bibnamefont {Janssen}},
  \bibinfo {author} {\bibfnamefont {P.}~\bibnamefont {Lampen-Kelley}}, \bibinfo
  {author} {\bibfnamefont {A.}~\bibnamefont {Banerjee}}, \bibinfo {author}
  {\bibfnamefont {Y.~H.}\ \bibnamefont {Liu}}, \bibinfo {author} {\bibfnamefont
  {J.-Q.}\ \bibnamefont {Yan}}, \bibinfo {author} {\bibfnamefont {D.~G.}\
  \bibnamefont {Mandrus}}, \bibinfo {author} {\bibfnamefont {M.}~\bibnamefont
  {Vojta}},\ and\ \bibinfo {author} {\bibfnamefont {S.~E.}\ \bibnamefont
  {Nagler}},\ }\bibfield  {title} {\bibinfo {title} {Field-induced intermediate
  ordered phase and anisotropic interlayer interactions in
  $\ensuremath{\alpha}\text{\ensuremath{-}}{\mathrm{rucl}}_{3}$},\ }\href
  {https://doi.org/10.1103/PhysRevB.103.174417} {\bibfield  {journal} {\bibinfo
   {journal} {Phys. Rev. B}\ }\textbf {\bibinfo {volume} {103}},\ \bibinfo
  {pages} {174417} (\bibinfo {year} {2021})}\BibitemShut {NoStop}%
\bibitem [{\citenamefont {Keimer}\ \emph {et~al.}(1992)\citenamefont {Keimer},
  \citenamefont {Aharony}, \citenamefont {Auerbach}, \citenamefont {Birgeneau},
  \citenamefont {Cassanho}, \citenamefont {Endoh}, \citenamefont {Erwin},
  \citenamefont {Kastner},\ and\ \citenamefont {Shirane}}]{PhysRevB.45.7430}%
  \BibitemOpen
  \bibfield  {author} {\bibinfo {author} {\bibfnamefont {B.}~\bibnamefont
  {Keimer}}, \bibinfo {author} {\bibfnamefont {A.}~\bibnamefont {Aharony}},
  \bibinfo {author} {\bibfnamefont {A.}~\bibnamefont {Auerbach}}, \bibinfo
  {author} {\bibfnamefont {R.~J.}\ \bibnamefont {Birgeneau}}, \bibinfo {author}
  {\bibfnamefont {A.}~\bibnamefont {Cassanho}}, \bibinfo {author}
  {\bibfnamefont {Y.}~\bibnamefont {Endoh}}, \bibinfo {author} {\bibfnamefont
  {R.~W.}\ \bibnamefont {Erwin}}, \bibinfo {author} {\bibfnamefont {M.~A.}\
  \bibnamefont {Kastner}},\ and\ \bibinfo {author} {\bibfnamefont
  {G.}~\bibnamefont {Shirane}},\ }\bibfield  {title} {\bibinfo {title} {N\'eel
  transition and sublattice magnetization of pure and doped
  ${\mathrm{la}}_{2}$${\mathrm{cuo}}_{4}$},\ }\href
  {https://doi.org/10.1103/PhysRevB.45.7430} {\bibfield  {journal} {\bibinfo
  {journal} {Phys. Rev. B}\ }\textbf {\bibinfo {volume} {45}},\ \bibinfo
  {pages} {7430} (\bibinfo {year} {1992})}\BibitemShut {NoStop}%
\bibitem [{\citenamefont {Kim}\ and\ \citenamefont
  {Kee}(2016)}]{PhysRevB.93.155143}%
  \BibitemOpen
  \bibfield  {author} {\bibinfo {author} {\bibfnamefont {H.-S.}\ \bibnamefont
  {Kim}}\ and\ \bibinfo {author} {\bibfnamefont {H.-Y.}\ \bibnamefont {Kee}},\
  }\bibfield  {title} {\bibinfo {title} {Crystal structure and magnetism in
  $\ensuremath{\alpha}\ensuremath{-}{\mathrm{rucl}}_{3}$: An ab initio study},\
  }\href {https://doi.org/10.1103/PhysRevB.93.155143} {\bibfield  {journal}
  {\bibinfo  {journal} {Phys. Rev. B}\ }\textbf {\bibinfo {volume} {93}},\
  \bibinfo {pages} {155143} (\bibinfo {year} {2016})}\BibitemShut {NoStop}%
\bibitem [{\citenamefont {Slagle}\ \emph {et~al.}(2018)\citenamefont {Slagle},
  \citenamefont {Choi}, \citenamefont {Chern},\ and\ \citenamefont
  {Kim}}]{slagle18}%
  \BibitemOpen
  \bibfield  {author} {\bibinfo {author} {\bibfnamefont {K.}~\bibnamefont
  {Slagle}}, \bibinfo {author} {\bibfnamefont {W.}~\bibnamefont {Choi}},
  \bibinfo {author} {\bibfnamefont {L.~E.}\ \bibnamefont {Chern}},\ and\
  \bibinfo {author} {\bibfnamefont {Y.~B.}\ \bibnamefont {Kim}},\ }\bibfield
  {title} {\bibinfo {title} {Theory of a quantum spin liquid in the
  hydrogen-intercalated honeycomb iridate
  ${\mathrm{h}}_{3}{\mathrm{liir}}_{2}{\mathrm{o}}_{6}$},\ }\href
  {https://doi.org/10.1103/PhysRevB.97.115159} {\bibfield  {journal} {\bibinfo
  {journal} {Phys. Rev. B}\ }\textbf {\bibinfo {volume} {97}},\ \bibinfo
  {pages} {115159} (\bibinfo {year} {2018})}\BibitemShut {NoStop}%
\bibitem [{\citenamefont {Cao}\ \emph {et~al.}(2016)\citenamefont {Cao},
  \citenamefont {Banerjee}, \citenamefont {Yan}, \citenamefont {Bridges},
  \citenamefont {Lumsden}, \citenamefont {Mandrus}, \citenamefont {Tennant},
  \citenamefont {Chakoumakos},\ and\ \citenamefont
  {Nagler}}]{PhysRevB.93.134423}%
  \BibitemOpen
  \bibfield  {author} {\bibinfo {author} {\bibfnamefont {H.~B.}\ \bibnamefont
  {Cao}}, \bibinfo {author} {\bibfnamefont {A.}~\bibnamefont {Banerjee}},
  \bibinfo {author} {\bibfnamefont {J.-Q.}\ \bibnamefont {Yan}}, \bibinfo
  {author} {\bibfnamefont {C.~A.}\ \bibnamefont {Bridges}}, \bibinfo {author}
  {\bibfnamefont {M.~D.}\ \bibnamefont {Lumsden}}, \bibinfo {author}
  {\bibfnamefont {D.~G.}\ \bibnamefont {Mandrus}}, \bibinfo {author}
  {\bibfnamefont {D.~A.}\ \bibnamefont {Tennant}}, \bibinfo {author}
  {\bibfnamefont {B.~C.}\ \bibnamefont {Chakoumakos}},\ and\ \bibinfo {author}
  {\bibfnamefont {S.~E.}\ \bibnamefont {Nagler}},\ }\bibfield  {title}
  {\bibinfo {title} {Low-temperature crystal and magnetic structure of
  $\ensuremath{\alpha}\ensuremath{-}{\mathrm{rucl}}_{3}$},\ }\href
  {https://doi.org/10.1103/PhysRevB.93.134423} {\bibfield  {journal} {\bibinfo
  {journal} {Phys. Rev. B}\ }\textbf {\bibinfo {volume} {93}},\ \bibinfo
  {pages} {134423} (\bibinfo {year} {2016})}\BibitemShut {NoStop}%
\bibitem [{\citenamefont {Werman}\ \emph {et~al.}(2018)\citenamefont {Werman},
  \citenamefont {Chatterjee}, \citenamefont {Morampudi},\ and\ \citenamefont
  {Berg}}]{WermanPRX}%
  \BibitemOpen
  \bibfield  {author} {\bibinfo {author} {\bibfnamefont {Y.}~\bibnamefont
  {Werman}}, \bibinfo {author} {\bibfnamefont {S.}~\bibnamefont {Chatterjee}},
  \bibinfo {author} {\bibfnamefont {S.~C.}\ \bibnamefont {Morampudi}},\ and\
  \bibinfo {author} {\bibfnamefont {E.}~\bibnamefont {Berg}},\ }\bibfield
  {title} {\bibinfo {title} {Signatures of fractionalization in spin liquids
  from interlayer thermal transport},\ }\href
  {https://doi.org/10.1103/PhysRevX.8.031064} {\bibfield  {journal} {\bibinfo
  {journal} {Phys. Rev. X}\ }\textbf {\bibinfo {volume} {8}},\ \bibinfo {pages}
  {031064} (\bibinfo {year} {2018})}\BibitemShut {NoStop}%
\bibitem [{\citenamefont {Tomishige}\ \emph {et~al.}(2019)\citenamefont
  {Tomishige}, \citenamefont {Nasu},\ and\ \citenamefont {Koga}}]{Tomishige19}%
  \BibitemOpen
  \bibfield  {author} {\bibinfo {author} {\bibfnamefont {H.}~\bibnamefont
  {Tomishige}}, \bibinfo {author} {\bibfnamefont {J.}~\bibnamefont {Nasu}},\
  and\ \bibinfo {author} {\bibfnamefont {A.}~\bibnamefont {Koga}},\ }\bibfield
  {title} {\bibinfo {title} {Low-temperature properties in the bilayer kitaev
  model},\ }\href {https://doi.org/10.1103/PhysRevB.99.174424} {\bibfield
  {journal} {\bibinfo  {journal} {Phys. Rev. B}\ }\textbf {\bibinfo {volume}
  {99}},\ \bibinfo {pages} {174424} (\bibinfo {year} {2019})}\BibitemShut
  {NoStop}%
\bibitem [{\citenamefont {Seifert}\ \emph {et~al.}(2018)\citenamefont
  {Seifert}, \citenamefont {Gritsch}, \citenamefont {Wagner}, \citenamefont
  {Joshi}, \citenamefont {Brenig}, \citenamefont {Vojta},\ and\ \citenamefont
  {Schmidt}}]{Urbanbilayer1}%
  \BibitemOpen
  \bibfield  {author} {\bibinfo {author} {\bibfnamefont {U.~F.~P.}\
  \bibnamefont {Seifert}}, \bibinfo {author} {\bibfnamefont {J.}~\bibnamefont
  {Gritsch}}, \bibinfo {author} {\bibfnamefont {E.}~\bibnamefont {Wagner}},
  \bibinfo {author} {\bibfnamefont {D.~G.}\ \bibnamefont {Joshi}}, \bibinfo
  {author} {\bibfnamefont {W.}~\bibnamefont {Brenig}}, \bibinfo {author}
  {\bibfnamefont {M.}~\bibnamefont {Vojta}},\ and\ \bibinfo {author}
  {\bibfnamefont {K.~P.}\ \bibnamefont {Schmidt}},\ }\bibfield  {title}
  {\bibinfo {title} {Bilayer kitaev models: Phase diagrams and novel phases},\
  }\href {https://doi.org/10.1103/PhysRevB.98.155101} {\bibfield  {journal}
  {\bibinfo  {journal} {Phys. Rev. B}\ }\textbf {\bibinfo {volume} {98}},\
  \bibinfo {pages} {155101} (\bibinfo {year} {2018})}\BibitemShut {NoStop}%
\bibitem [{\citenamefont {Vijayvargia}\ \emph {et~al.}(2024)\citenamefont
  {Vijayvargia}, \citenamefont {Seifert},\ and\ \citenamefont
  {Erten}}]{Urbanbilayer2}%
  \BibitemOpen
  \bibfield  {author} {\bibinfo {author} {\bibfnamefont {A.}~\bibnamefont
  {Vijayvargia}}, \bibinfo {author} {\bibfnamefont {U.~F.~P.}\ \bibnamefont
  {Seifert}},\ and\ \bibinfo {author} {\bibfnamefont {O.}~\bibnamefont
  {Erten}},\ }\bibfield  {title} {\bibinfo {title} {Topological and magnetic
  phase transitions in the bilayer kitaev-ising model},\ }\href
  {https://doi.org/10.1103/PhysRevB.109.024439} {\bibfield  {journal} {\bibinfo
   {journal} {Phys. Rev. B}\ }\textbf {\bibinfo {volume} {109}},\ \bibinfo
  {pages} {024439} (\bibinfo {year} {2024})}\BibitemShut {NoStop}%
\bibitem [{\citenamefont {Wiedmann}\ \emph {et~al.}(2020)\citenamefont
  {Wiedmann}, \citenamefont {Lenke}, \citenamefont {Walther}, \citenamefont
  {M\"uhlhauser},\ and\ \citenamefont {Schmidt}}]{tcbilayer}%
  \BibitemOpen
  \bibfield  {author} {\bibinfo {author} {\bibfnamefont {R.}~\bibnamefont
  {Wiedmann}}, \bibinfo {author} {\bibfnamefont {L.}~\bibnamefont {Lenke}},
  \bibinfo {author} {\bibfnamefont {M.~R.}\ \bibnamefont {Walther}}, \bibinfo
  {author} {\bibfnamefont {M.}~\bibnamefont {M\"uhlhauser}},\ and\ \bibinfo
  {author} {\bibfnamefont {K.~P.}\ \bibnamefont {Schmidt}},\ }\bibfield
  {title} {\bibinfo {title} {Quantum critical phase transition between two
  topologically ordered phases in the ising toric code bilayer},\ }\href
  {https://doi.org/10.1103/PhysRevB.102.214422} {\bibfield  {journal} {\bibinfo
   {journal} {Phys. Rev. B}\ }\textbf {\bibinfo {volume} {102}},\ \bibinfo
  {pages} {214422} (\bibinfo {year} {2020})}\BibitemShut {NoStop}%
\bibitem [{\citenamefont {Ma}\ \emph {et~al.}(2017)\citenamefont {Ma},
  \citenamefont {Lake}, \citenamefont {Chen},\ and\ \citenamefont
  {Hermele}}]{Ma17}%
  \BibitemOpen
  \bibfield  {author} {\bibinfo {author} {\bibfnamefont {H.}~\bibnamefont
  {Ma}}, \bibinfo {author} {\bibfnamefont {E.}~\bibnamefont {Lake}}, \bibinfo
  {author} {\bibfnamefont {X.}~\bibnamefont {Chen}},\ and\ \bibinfo {author}
  {\bibfnamefont {M.}~\bibnamefont {Hermele}},\ }\bibfield  {title} {\bibinfo
  {title} {Fracton topological order via coupled layers},\ }\href
  {https://doi.org/10.1103/PhysRevB.95.245126} {\bibfield  {journal} {\bibinfo
  {journal} {Phys. Rev. B}\ }\textbf {\bibinfo {volume} {95}},\ \bibinfo
  {pages} {245126} (\bibinfo {year} {2017})}\BibitemShut {NoStop}%
\bibitem [{\citenamefont {Slagle}\ and\ \citenamefont {Kim}(2017)}]{slagle17}%
  \BibitemOpen
  \bibfield  {author} {\bibinfo {author} {\bibfnamefont {K.}~\bibnamefont
  {Slagle}}\ and\ \bibinfo {author} {\bibfnamefont {Y.~B.}\ \bibnamefont
  {Kim}},\ }\bibfield  {title} {\bibinfo {title} {Fracton topological order
  from nearest-neighbor two-spin interactions and dualities},\ }\href
  {https://doi.org/10.1103/PhysRevB.96.165106} {\bibfield  {journal} {\bibinfo
  {journal} {Phys. Rev. B}\ }\textbf {\bibinfo {volume} {96}},\ \bibinfo
  {pages} {165106} (\bibinfo {year} {2017})}\BibitemShut {NoStop}%
\bibitem [{\citenamefont {Schamri\ss{}}\ \emph {et~al.}(2022)\citenamefont
  {Schamri\ss{}}, \citenamefont {Lenke}, \citenamefont {M\"uhlhauser},\ and\
  \citenamefont {Schmidt}}]{KLayerToricCode}%
  \BibitemOpen
  \bibfield  {author} {\bibinfo {author} {\bibfnamefont {L.}~\bibnamefont
  {Schamri\ss{}}}, \bibinfo {author} {\bibfnamefont {L.}~\bibnamefont {Lenke}},
  \bibinfo {author} {\bibfnamefont {M.}~\bibnamefont {M\"uhlhauser}},\ and\
  \bibinfo {author} {\bibfnamefont {K.~P.}\ \bibnamefont {Schmidt}},\
  }\bibfield  {title} {\bibinfo {title} {Quantum phase transitions in the
  $k$-layer ising toric code},\ }\href
  {https://doi.org/10.1103/PhysRevB.105.184425} {\bibfield  {journal} {\bibinfo
   {journal} {Phys. Rev. B}\ }\textbf {\bibinfo {volume} {105}},\ \bibinfo
  {pages} {184425} (\bibinfo {year} {2022})}\BibitemShut {NoStop}%
\bibitem [{\citenamefont {Robledo}(2009)}]{robledo}%
  \BibitemOpen
  \bibfield  {author} {\bibinfo {author} {\bibfnamefont {L.~M.}\ \bibnamefont
  {Robledo}},\ }\bibfield  {title} {\bibinfo {title} {Sign of the overlap of
  hartree-fock-bogoliubov wave functions},\ }\href
  {https://doi.org/10.1103/PhysRevC.79.021302} {\bibfield  {journal} {\bibinfo
  {journal} {Phys. Rev. C}\ }\textbf {\bibinfo {volume} {79}},\ \bibinfo
  {pages} {021302} (\bibinfo {year} {2009})}\BibitemShut {NoStop}%
\bibitem [{\citenamefont {Knolle}\ \emph {et~al.}(2014)\citenamefont {Knolle},
  \citenamefont {Kovrizhin}, \citenamefont {Chalker},\ and\ \citenamefont
  {Moessner}}]{knolle}%
  \BibitemOpen
  \bibfield  {author} {\bibinfo {author} {\bibfnamefont {J.}~\bibnamefont
  {Knolle}}, \bibinfo {author} {\bibfnamefont {D.~L.}\ \bibnamefont
  {Kovrizhin}}, \bibinfo {author} {\bibfnamefont {J.~T.}\ \bibnamefont
  {Chalker}},\ and\ \bibinfo {author} {\bibfnamefont {R.}~\bibnamefont
  {Moessner}},\ }\bibfield  {title} {\bibinfo {title} {{Dynamics of a
  Two-Dimensional Quantum Spin Liquid: Signatures of Emergent Majorana Fermions
  and Fluxes}},\ }\href {https://doi.org/10.1103/PhysRevLett.112.207203}
  {\bibfield  {journal} {\bibinfo  {journal} {Phys. Rev. Lett.}\ }\textbf
  {\bibinfo {volume} {112}},\ \bibinfo {pages} {207203} (\bibinfo {year}
  {2014})}\BibitemShut {NoStop}%
\bibitem [{\citenamefont {Pedrocchi}\ \emph {et~al.}(2011)\citenamefont
  {Pedrocchi}, \citenamefont {Chesi},\ and\ \citenamefont {Loss}}]{Loss}%
  \BibitemOpen
  \bibfield  {author} {\bibinfo {author} {\bibfnamefont {F.~L.}\ \bibnamefont
  {Pedrocchi}}, \bibinfo {author} {\bibfnamefont {S.}~\bibnamefont {Chesi}},\
  and\ \bibinfo {author} {\bibfnamefont {D.}~\bibnamefont {Loss}},\ }\bibfield
  {title} {\bibinfo {title} {{Physical solutions of the Kitaev honeycomb
  model}},\ }\href {https://doi.org/10.1103/PhysRevB.84.165414} {\bibfield
  {journal} {\bibinfo  {journal} {Phys. Rev. B}\ }\textbf {\bibinfo {volume}
  {84}},\ \bibinfo {pages} {165414} (\bibinfo {year} {2011})}\BibitemShut
  {NoStop}%
\bibitem [{\citenamefont {Vojta}\ and\ \citenamefont {Zschocke}(2015)}]{vojta}%
  \BibitemOpen
  \bibfield  {author} {\bibinfo {author} {\bibfnamefont {M.}~\bibnamefont
  {Vojta}}\ and\ \bibinfo {author} {\bibfnamefont {F.}~\bibnamefont
  {Zschocke}},\ }\bibfield  {title} {\bibinfo {title} {{Physical states and
  finite-size effects in Kitaev's honeycomb model: Bond disorder, spin
  excitations, and NMR line shape}},\ }\href
  {https://link.aps.org/doi/10.1103/PhysRevB.92.014403} {\bibfield  {journal}
  {\bibinfo  {journal} {Physical Review B}\ }\textbf {\bibinfo {volume} {92}}
  (\bibinfo {year} {2015})}\BibitemShut {NoStop}%
\bibitem [{\citenamefont {Panigrahi}\ \emph {et~al.}(2023)\citenamefont
  {Panigrahi}, \citenamefont {Coleman},\ and\ \citenamefont
  {Tsvelik}}]{aadityaPRB}%
  \BibitemOpen
  \bibfield  {author} {\bibinfo {author} {\bibfnamefont {A.}~\bibnamefont
  {Panigrahi}}, \bibinfo {author} {\bibfnamefont {P.}~\bibnamefont {Coleman}},\
  and\ \bibinfo {author} {\bibfnamefont {A.}~\bibnamefont {Tsvelik}},\
  }\bibfield  {title} {\bibinfo {title} {Analytic calculation of the vison gap
  in the kitaev spin liquid},\ }\href
  {https://doi.org/10.1103/PhysRevB.108.045151} {\bibfield  {journal} {\bibinfo
   {journal} {Phys. Rev. B}\ }\textbf {\bibinfo {volume} {108}},\ \bibinfo
  {pages} {045151} (\bibinfo {year} {2023})}\BibitemShut {NoStop}%
\bibitem [{\citenamefont {Balz}\ \emph {et~al.}(2019)\citenamefont {Balz},
  \citenamefont {Lampen-Kelley}, \citenamefont {Banerjee}, \citenamefont {Yan},
  \citenamefont {Lu}, \citenamefont {Hu}, \citenamefont {Yadav}, \citenamefont
  {Takano}, \citenamefont {Liu}, \citenamefont {Tennant}, \citenamefont
  {Lumsden}, \citenamefont {Mandrus},\ and\ \citenamefont {Nagler}}]{Balz19}%
  \BibitemOpen
  \bibfield  {author} {\bibinfo {author} {\bibfnamefont {C.}~\bibnamefont
  {Balz}}, \bibinfo {author} {\bibfnamefont {P.}~\bibnamefont {Lampen-Kelley}},
  \bibinfo {author} {\bibfnamefont {A.}~\bibnamefont {Banerjee}}, \bibinfo
  {author} {\bibfnamefont {J.}~\bibnamefont {Yan}}, \bibinfo {author}
  {\bibfnamefont {Z.}~\bibnamefont {Lu}}, \bibinfo {author} {\bibfnamefont
  {X.}~\bibnamefont {Hu}}, \bibinfo {author} {\bibfnamefont {S.~M.}\
  \bibnamefont {Yadav}}, \bibinfo {author} {\bibfnamefont {Y.}~\bibnamefont
  {Takano}}, \bibinfo {author} {\bibfnamefont {Y.}~\bibnamefont {Liu}},
  \bibinfo {author} {\bibfnamefont {D.~A.}\ \bibnamefont {Tennant}}, \bibinfo
  {author} {\bibfnamefont {M.~D.}\ \bibnamefont {Lumsden}}, \bibinfo {author}
  {\bibfnamefont {D.}~\bibnamefont {Mandrus}},\ and\ \bibinfo {author}
  {\bibfnamefont {S.~E.}\ \bibnamefont {Nagler}},\ }\bibfield  {title}
  {\bibinfo {title} {Finite field regime for a quantum spin liquid in
  $\ensuremath{\alpha}\text{\ensuremath{-}}{\mathrm{rucl}}_{3}$},\ }\href
  {https://doi.org/10.1103/PhysRevB.100.060405} {\bibfield  {journal} {\bibinfo
   {journal} {Phys. Rev. B}\ }\textbf {\bibinfo {volume} {100}},\ \bibinfo
  {pages} {060405} (\bibinfo {year} {2019})}\BibitemShut {NoStop}%
\bibitem [{\citenamefont {Janssen}\ \emph {et~al.}(2020)\citenamefont
  {Janssen}, \citenamefont {Koch},\ and\ \citenamefont {Vojta}}]{Jansen20}%
  \BibitemOpen
  \bibfield  {author} {\bibinfo {author} {\bibfnamefont {L.}~\bibnamefont
  {Janssen}}, \bibinfo {author} {\bibfnamefont {S.}~\bibnamefont {Koch}},\ and\
  \bibinfo {author} {\bibfnamefont {M.}~\bibnamefont {Vojta}},\ }\bibfield
  {title} {\bibinfo {title} {Magnon dispersion and dynamic spin response in
  three-dimensional spin models for
  $\ensuremath{\alpha}\text{\ensuremath{-}}{\mathrm{rucl}}_{3}$},\ }\href
  {https://doi.org/10.1103/PhysRevB.101.174444} {\bibfield  {journal} {\bibinfo
   {journal} {Phys. Rev. B}\ }\textbf {\bibinfo {volume} {101}},\ \bibinfo
  {pages} {174444} (\bibinfo {year} {2020})}\BibitemShut {NoStop}%
\bibitem [{\citenamefont {Gromov}\ and\ \citenamefont
  {Radzihovsky}(2024)}]{fractonreview}%
  \BibitemOpen
  \bibfield  {author} {\bibinfo {author} {\bibfnamefont {A.}~\bibnamefont
  {Gromov}}\ and\ \bibinfo {author} {\bibfnamefont {L.}~\bibnamefont
  {Radzihovsky}},\ }\bibfield  {title} {\bibinfo {title} {Colloquium: Fracton
  matter},\ }\href {https://doi.org/10.1103/RevModPhys.96.011001} {\bibfield
  {journal} {\bibinfo  {journal} {Rev. Mod. Phys.}\ }\textbf {\bibinfo {volume}
  {96}},\ \bibinfo {pages} {011001} (\bibinfo {year} {2024})}\BibitemShut
  {NoStop}%
\bibitem [{\citenamefont {Pretko}\ \emph {et~al.}(2020)\citenamefont {Pretko},
  \citenamefont {Chen},\ and\ \citenamefont {You}}]{fractonPretko}%
  \BibitemOpen
  \bibfield  {author} {\bibinfo {author} {\bibfnamefont {M.}~\bibnamefont
  {Pretko}}, \bibinfo {author} {\bibfnamefont {X.}~\bibnamefont {Chen}},\ and\
  \bibinfo {author} {\bibfnamefont {Y.}~\bibnamefont {You}},\ }\bibfield
  {title} {\bibinfo {title} {Fracton phases of matter},\ }\href
  {https://doi.org/10.1142/S0217751X20300033} {\bibfield  {journal} {\bibinfo
  {journal} {International Journal of Modern Physics A}\ }\textbf {\bibinfo
  {volume} {35}},\ \bibinfo {pages} {2030003} (\bibinfo {year} {2020})},\
  \Eprint {https://arxiv.org/abs/https://doi.org/10.1142/S0217751X20300033}
  {https://doi.org/10.1142/S0217751X20300033} \BibitemShut {NoStop}%
\bibitem [{\citenamefont {Hwang}(2024)}]{kyusungPRB}%
  \BibitemOpen
  \bibfield  {author} {\bibinfo {author} {\bibfnamefont {K.}~\bibnamefont
  {Hwang}},\ }\bibfield  {title} {\bibinfo {title} {Anyon condensation and
  confinement transition in a kitaev spin liquid bilayer},\ }\href
  {https://doi.org/10.1103/PhysRevB.109.134412} {\bibfield  {journal} {\bibinfo
   {journal} {Phys. Rev. B}\ }\textbf {\bibinfo {volume} {109}},\ \bibinfo
  {pages} {134412} (\bibinfo {year} {2024})}\BibitemShut {NoStop}%
\bibitem [{\citenamefont {Kohno}\ \emph {et~al.}(2007)\citenamefont {Kohno},
  \citenamefont {Starykh},\ and\ \citenamefont {Balents}}]{kohno2007spinons}%
  \BibitemOpen
  \bibfield  {author} {\bibinfo {author} {\bibfnamefont {M.}~\bibnamefont
  {Kohno}}, \bibinfo {author} {\bibfnamefont {O.~A.}\ \bibnamefont {Starykh}},\
  and\ \bibinfo {author} {\bibfnamefont {L.}~\bibnamefont {Balents}},\
  }\bibfield  {title} {\bibinfo {title} {Spinons and triplons in spatially
  anisotropic frustrated antiferromagnets},\ }\href@noop {} {\bibfield
  {journal} {\bibinfo  {journal} {Nature Physics}\ }\textbf {\bibinfo {volume}
  {3}},\ \bibinfo {pages} {790} (\bibinfo {year} {2007})}\BibitemShut {NoStop}%
\bibitem [{\citenamefont {Versteeg}\ \emph {et~al.}(2022)\citenamefont
  {Versteeg}, \citenamefont {Chiocchetta}, \citenamefont {Sekiguchi},
  \citenamefont {Sahasrabudhe}, \citenamefont {Wagner}, \citenamefont {Aldea},
  \citenamefont {Budzinauskas}, \citenamefont {Wang}, \citenamefont {Tsurkan},
  \citenamefont {Loidl}, \citenamefont {Khomskii}, \citenamefont {Diehl},\ and\
  \citenamefont {van Loosdrecht}}]{versteeg22}%
  \BibitemOpen
  \bibfield  {author} {\bibinfo {author} {\bibfnamefont {R.~B.}\ \bibnamefont
  {Versteeg}}, \bibinfo {author} {\bibfnamefont {A.}~\bibnamefont
  {Chiocchetta}}, \bibinfo {author} {\bibfnamefont {F.}~\bibnamefont
  {Sekiguchi}}, \bibinfo {author} {\bibfnamefont {A.}~\bibnamefont
  {Sahasrabudhe}}, \bibinfo {author} {\bibfnamefont {J.}~\bibnamefont
  {Wagner}}, \bibinfo {author} {\bibfnamefont {A.~I.~R.}\ \bibnamefont
  {Aldea}}, \bibinfo {author} {\bibfnamefont {K.}~\bibnamefont {Budzinauskas}},
  \bibinfo {author} {\bibfnamefont {Z.}~\bibnamefont {Wang}}, \bibinfo {author}
  {\bibfnamefont {V.}~\bibnamefont {Tsurkan}}, \bibinfo {author} {\bibfnamefont
  {A.}~\bibnamefont {Loidl}}, \bibinfo {author} {\bibfnamefont {D.~I.}\
  \bibnamefont {Khomskii}}, \bibinfo {author} {\bibfnamefont {S.}~\bibnamefont
  {Diehl}},\ and\ \bibinfo {author} {\bibfnamefont {P.~H.~M.}\ \bibnamefont
  {van Loosdrecht}},\ }\bibfield  {title} {\bibinfo {title} {Nonequilibrium
  quasistationary spin disordered state in
  $\ensuremath{\alpha}\text{\ensuremath{-}}{\mathrm{rucl}}_{3}$},\ }\href
  {https://doi.org/10.1103/PhysRevB.105.224428} {\bibfield  {journal} {\bibinfo
   {journal} {Phys. Rev. B}\ }\textbf {\bibinfo {volume} {105}},\ \bibinfo
  {pages} {224428} (\bibinfo {year} {2022})}\BibitemShut {NoStop}%
\bibitem [{\citenamefont {Joy}(2024)}]{dataZenodo}%
  \BibitemOpen
  \bibfield  {author} {\bibinfo {author} {\bibfnamefont {A.}~\bibnamefont
  {Joy}},\ }\bibfield  {title} {\bibinfo {title} {{Data generated from
  calculations in "Gauge Field Dynamics in a Multilayer Kitaev Spin Liquid"}},\
  }\href {https://doi.org/10.5281/zenodo.10876207} {10.5281/zenodo.10876207}
  (\bibinfo {year} {2024})\BibitemShut {NoStop}%
\end{thebibliography}%

\

\end{document}